\documentclass[fleqn,usenatbib]{mnras}
\usepackage{newtxtext,newtxmath}
\usepackage[T1]{fontenc}
\usepackage{ae,aecompl}

\usepackage{graphicx}	% Including figure files
\usepackage{amsmath}	% Advanced maths commands
\usepackage{amssymb}	% Extra maths symbols
\usepackage{longtable}
\usepackage[usenames, dvipsnames]{color}
\usepackage{times}
\usepackage{subfigure}
\usepackage{epsfig}
\usepackage{graphics}
\usepackage{multicol}
\usepackage[toc,page]{appendix} 
\usepackage{hyperref}

\renewcommand{\descriptionlabel}[1]%
  {\hspace{\labelsep}\textbf{#1}}

\title[NGC 6712: the variable star population]{NGC 6712: the variable star population of a tidally disrupted globular cluster.$\thanks{Based on observations made with the telescope IAC80, in the Spanish Observatorio del Teide of the Instituto de Astrof\'isica de Canarias, in the island of Tenerife, Spain, and with the 2.0 m telescope at the Indian Astrophysical Observatory,
Hanle, India.}$}

\author[D. Deras et al.]{
D. Deras,$^{1}$\thanks{E-mail: dderas@astro.unam.mx, 
armando@astro.unam.mx, clh@iac.es, ivanbf@oac.unc.edu.ar,
calderon@oac.unc.edu.ar, smuneer@iiap.res.in, giridhar@iiap.res.in}
A. Arellano Ferro$^{1}$, C. L\'azaro$^{2,3}$, I.H. Bustos Fierro$^{4}$, J. H. Calder\'on$^{4,5}$, \and
S. Muneer$^{6}$, Sunetra Giridhar$^{6}$\\
$^{1}$Instituto de Astronom\'ia, Universidad Nacional Aut\'onoma de M\'exico, Ciudad Universitaria, C.P. 04510, M\'exico.\\	
$^{2}$Departamento de Astrof\'isica, Universidad de La Laguna, E-38206 La Laguna, Tenerife, Spain.\\
$^{3}$Instituto de Astrof\'isica de Canarias (IAC), E-38205 La Laguna, Tenerife,
Spain.\\
$^{4}$Observatorio Astron\'omico, Universidad Nacional
de C\'ordoba, C\'ordoba, Argentina.\\
$^{5}$Consejo Nacional de Investigaciones Cient\'ificas y
T\'ecnicas (CONICET), Buenos Aires, Argentina.\\
$^{6}$Indian Institute of Astrophysics, Koramangala 560034, Bangalore, India.\\
}

\date{Accepted XXX. Received YYY; in original form ZZZ}

\pubyear{2019}
\begin{document}
\label{firstpage}
\pagerange{\pageref{firstpage}--\pageref{lastpage}}
\maketitle

\begin{abstract}

We present an analysis of $\emph{VI}$ CCD time series photometry of globular cluster NGC 6712. Our main goal is to study the variable star population as indicators of the cluster mean physical parameters. We employed the Fourier decomposition of RR Lyrae light curves to confirm that $[\rm Fe/H]_{UVES} = -1.0 \pm 0.05$ is a solid estimate.  We estimated the reddening to the cluster as $E(B-V)$ = 0.35 $\pm$ 0.04 from the RRab stars colour curves. The distance to the cluster was estimated using three independent methods which yielded a weighted mean distance $\big < d \big>$ = 8.1 $\pm$ 0.2 kpc. The distribution of RRab and RRc stars on the HB shows a clear segregation around the first overtone red edge of the instability strip, which seems to be a common feature in OoI-type cluster with a very red horizontal branch. We carried out a membership analysis of 60,447 stars in our FoV using the data from $Gaia$-DR2 and found 1529 likely members; we possess the light curves of 1100 among the member stars. This allowed us to produce a clean colour-magnitude diagram, consistent with an age of 12 Gyrs, and enabled us to discover close unresolved contaminants for several variable stars. From the proper motion analysis we found evidence of non-member stars in the FoV of the cluster being tidally affected by the gravitational pull of the bulge of the Galaxy. We found that the RRab variable V6, shows a previously undetected Blazhko effect. Finally, we report sixteen new variables of the EW-type (9) and SR-type (7).
\end{abstract}


%%%%%%%%%%%%%%%%%%%%%%%%%%%%%%%%%%%%%%%%%%%%%%%%%%

%%%%%%%%%%%%%%%%% BODY OF PAPER %%%%%%%%%%%%%%%%%%

\section{Introduction}
\label{sec:intro}
NGC 6712 is a small, sparse and metal rich globular cluster ($\rm [Fe/H]$ = $-1.02$; \citet{Harris1996}) that can be found behind the Scutum stellar cloud ($\alpha = 18^{h} 53' 04.32'', \delta = -08^{\circ} 42' 21.5''$, J2000), and is located in the populated region of the Galactic bulge ($l = 25.35^{\circ}$, $b = -4.32^{\circ}$). Two versions of the Galactic orbit of the cluster can be seen in Fig. 4. of \citet{Allen1990} and in Fig. 18 of \citet{Souza2017}.
The orbit is asymmetrical, it is confined to within 2.0 kpc from the Galactic disc and penetrates deep into the Galactic center, reaching a galactocentric distance of 0.2-0.3 kpc (\citealt{Dauphole1996}; \citealt{DeMarchi1999}). A recent estimation of the perigalacticon from \textit{Gaia}-DR2 proper motions is 0.45$\pm$0.10 kpc \citep{Baumgardt2019}.  According to  \citet{Cudworth1988}, the latest Galactic plane crossing could have happened as recently as $\sim$4000 years ago, which is much smaller than its half-mass relaxation time of 1 Gyr \citep{Harris1996}.
Hence, repeated visits to the bulge and constant interaction with the disc in its 12 Gyrs, makes it very likely that this cluster has experienced several and strong tidal disruptions (\citealt{Andreuzzi2001}; \citealt{Paltrinieri2001}). 
\newpage
Theoretical calculations of the mass function lead to the suspicion that the $\sim 10^5 M_{\odot}$ \citep{Pryor1993} cluster we see today is a remnant of the otherwise much more massive ($\sim 10^7 M_{\odot}$) cluster. This enormous mass loss may be reflected in the proper motions distribution, hence it is an added interest to explore the membership status of the stars in the field of the cluster and their motions in the light of the \textit{Gaia}-DR2 data.

The variable star population of the cluster includes RR Lyrae, long-period semi-regular variables, short-period eclipsing binaries and an X-ray source whose counterpart presents optical variability (\citealt{Pietrukowicz2004}; \citealt{Homer1996}). The Catalogue of Variable Stars in Globular Clusters (CVSGC) \citep{Clement2001}, in its 2015 edition, lists 28 confirmed variables and summarises the history of their discovery between 1917 and 2004. The potential of variable stars to estimate mean metallicity, reddening and distance for the globular cluster is well known. High quality CCD photometry of the stars in the field of view of the cluster provide insights, not only on the pulsating properties of individual variables and potential discovery of new ones, but also enables the discussion of the Colour-Magnitude diagram (CMD) structure and its relation with the stellar evolution patterns and age of the system. In the present study we propose to undertake a $\emph{VI}$ CCD time-series analysis of NGC 6712.
 
The paper is structured as follows: In $\S$  \ref{sec:data} we detail the observations and reduction process of our images, $\S$  \ref{membership} we perform a membership analysis of the stars in our FoV to the cluster based on the \textit{Gaia}-DR2 proper motions, in $\S$ \ref{variable_stars_6712} we describe our systematic search for variables and report new discoveries, in $\S$ \ref{decomposition} we describe the Fourier approach to estimate the physical parameters of the RR Lyrae stars, in $\S$ \ref{decomp_Nemec} we discuss the metallicity of the cluster employing alternative Fourier calibrations, in $\S$  \ref{distance_determination} we present the cluster distance estimations by several independent methods, in $\S$ \ref{CMD} we describe the structure of the CMD and the Horizontal Branch (HB), obtained from our photometry, in $\S$ \ref{dynamics} we briefly comment about the interaction of the cluster with the Galactic bulge and in $\S$  \ref{conclusions} we summarise our conclusions.

\section{Data, observations and reductions}
\label{sec:data}

The data used in the present work were obtained from two sites. The first set of data was obtained with the 0.80 m IAC80 telescope at the Observatorio del Teide (Tenerife, Spain), during 10 non-consecutive nights, between June 13th and July 16th of 2016. We used the CAMELOT camera with 2048x2048 pixels and 0.304 arcsec/pixel, with a 10.4$\times$10.4 arcmin$^2$ FoV, and a back illuminated detector CCD42-40 from E2V Technologies. 
The second set of data was obtained using the 2 m telescope at the Indian Astronomical Observatory (IAO) in Hanle, India on 12 nights separated into four intermittent seasons between June 2011 and April 2018. The detector used was a SITe ST-002 thinned backside illuminated CCD of 2048$\times$2048 pixels with a scale of 0.296 arcsec/pix, translating to a field of view (FoV) of approximately 10.1$\times$10.1~arcmin$^2$. The log of our observations is given in Table \ref{tab:observations}.
 
\begin{table}
\caption{The distribution of observations of NGC 6712 for each filter.
Columns $N_{V}$ and $N_{I}$ give the number of images taken with the $V$ and $I$
filters respectively. Columns $t_{V}$ and $t_{I}$ provide the exposure time,
or range of exposure times employed during each night for each filter. The 
average seeing is listed in the last column.}
\centering
\begin{tabular}{lcccccc}
\hline
Date &  $N_{V}$ & $t_{V}$ (s) & $N_{I}$ &$t_{I}$ (s) & Avg.& site$^1$ \\
 & && & & seeing (")& \\
\hline
20110610      & 8   & 70-100 & 8  & 10 & 1.5& H \\
20110611      & 7   & 70-150 & 10 & 15-40 & 2.1& H \\
20110805      & 15  & 120-150 & 16& 10-25 & 1.6& H \\
20110806      & 16  & 110-200 & 17 & 12-30 & 1.7& H \\
20130730      & 20   & 15-20 & 19  & 3-5 & 1.7& H \\
20150310      & 4   & 45 & 6  & 10 & 2.3& H \\
20150311      & 6   & 45 & 4  & 10 & 2.6& H \\
20150312      & 4   & 45 & 4  & 10 & 2.2& H \\
20150326      & 6   & 45 & 6  & 10 & 2.5& H \\
20150327      & 9   & 45-60 & 12  & 10-12 & 1.8& H \\
20160613      & 9   & 600-800 & 9  & 400-500& 1.6&I \\
20160615      & 23  & 80-100  & 22 & 20-30  & 2.0&I \\
20160626      & 21  & 80-100  & 23 & 20-50  & 2.0&I \\
20160627      & 89  & 40-80   & 88 & 10-20  & 1.2&I \\
20160628      & 2   & 40-50   & 3  & 15     & 2.0&I \\
20160629      & 40  & 20      & 45 & 60     & 2.8&I \\
20160702      & 100 & 40-60   & 99 & 10-20  & 1.5&I \\
20160703      & 94  & 20-80   & 90 & 5-20   & 1.2&I \\
20160704      & 78  & 13-60   & 74 & 10-20  & 1.7&I \\
20160716      & 24  & 50-80   & 24 & 13-30  & 1.8&I \\
20180408      & 4  & 45   & 6 & 15  & 2.2&H \\
20180408      & 5  & 20   & 6 & 7   & 2.4&H \\

\hline
Total:        & 584 & --      & 591 & -- & -- & -- \\
\hline
\end{tabular}
\center{1. H:Hanle, I:IAC }
\label{tab:observations}
\end{table}

\subsection{Difference Image Analysis}
The technique of Difference Imaging Analysis (DIA) with its pipeline implementation DanDIA \footnote{DanDIA is built from the DanIDL library of IDL routines
available at \url{http://www.danidl.co.uk}} (\citealt{Bramich2008}; \citealt{Bramich2013}) was used for the reduction of our data. This allowed us to obtain high-precision photometry for all the point sources in the FoV of the CCD. The software first creates a reference image by stacking the best images in each filter and then it subtracts them from the rest of the images in our collection. Using the PSF calculated by DanDIA from a sample of 300-400 isolated stars, we can determine the differential flux for each point source in the FoV. Then, the differential fluxes are converted into total fluxes.  
The total flux $f_{\mbox{\scriptsize tot}}(t)$ in ADU/s at each epoch $t$ can be estimated as:
\begin{equation}
f_{\mbox{\scriptsize tot}}(t) = f_{\mbox{\scriptsize ref}} +
\frac{f_{\mbox{\scriptsize diff}}(t)}{p(t)},
\label{eqn:totflux}
\end{equation}

\noindent
where $f_{\mbox{\scriptsize ref}}$ is the reference flux (ADU/s), $f_{\mbox{\scriptsize diff}}(t)$ is the differential flux (ADU/s) and
$p(t)$ is the photometric scale factor (the integral of the kernel solution). Conversion to instrumental magnitudes was achieved using:
\begin{equation}
m_{\mbox{\scriptsize ins}}(t) = 25.0 - 2.5 \log \left[ f_{\mbox{\scriptsize tot}}(t)
\right],
\label{eqn:mag}
\end{equation}

\noindent
where $m_{\mbox{\scriptsize ins}}(t)$ is the instrumental magnitude of the star at time $t$. A more detailed description of this method can be found in  \citet{Bramich2011}. 

Systematic errors in our DIA photometry may be introduced due to a possible error in the flux-magnitude conversion factor \citep{Bramich2015}. To investigate their significance, we applied the methodology developed in \citet{Bramich2012} to solve for the magnitude offsets $Z_{k}$ that should be applied to each photometric measurement from the image $k$. We found however, that in the present case the necessary corrections to our photometry were negligible (<0.001 mag).

\subsection{Transformation to the standard system}

\begin{figure}
\centerline{\includegraphics[width=8.5cm,height=5.5cm]{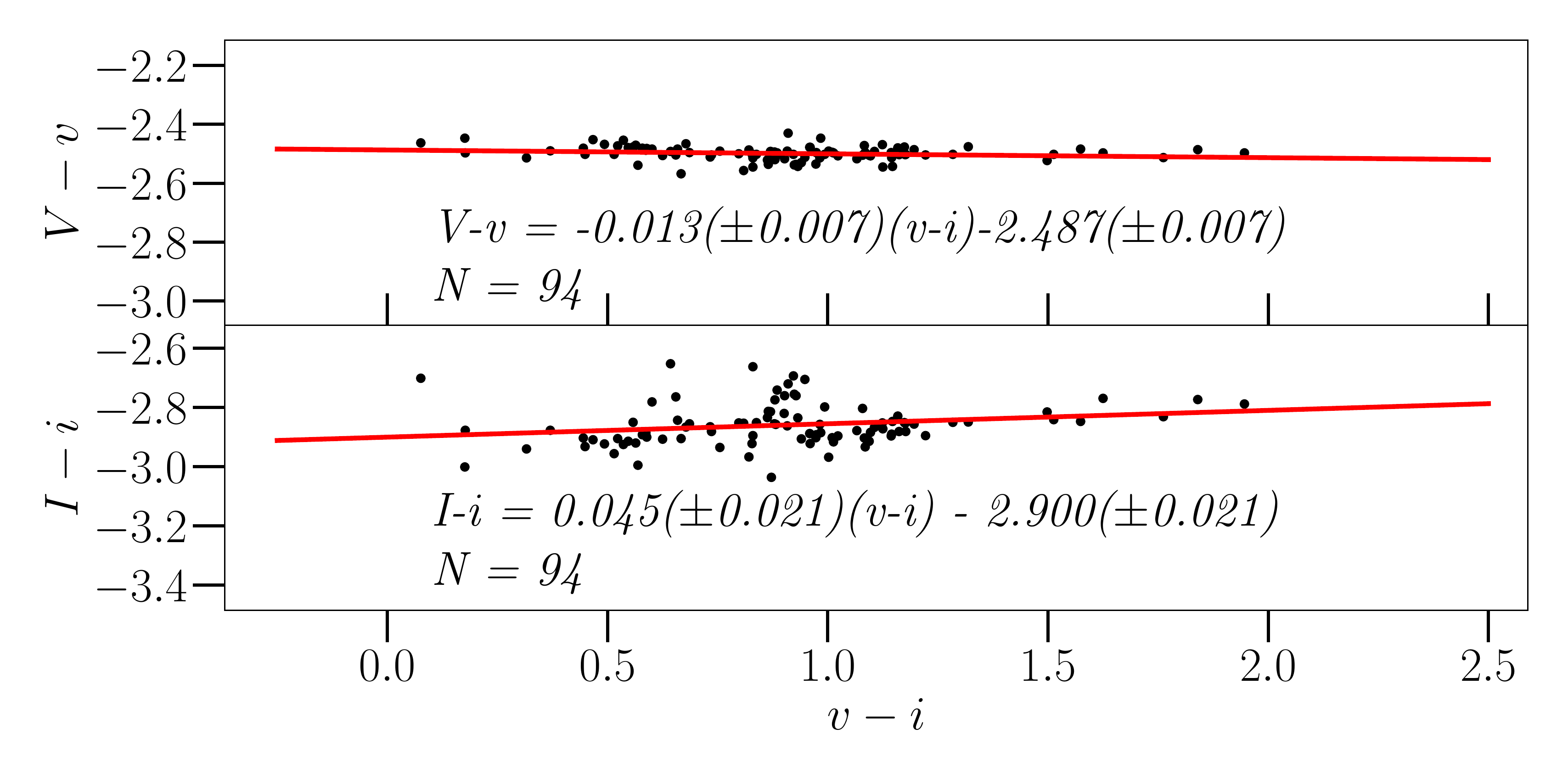}}

\centerline{\includegraphics[width=8.5cm,height=5.5cm]{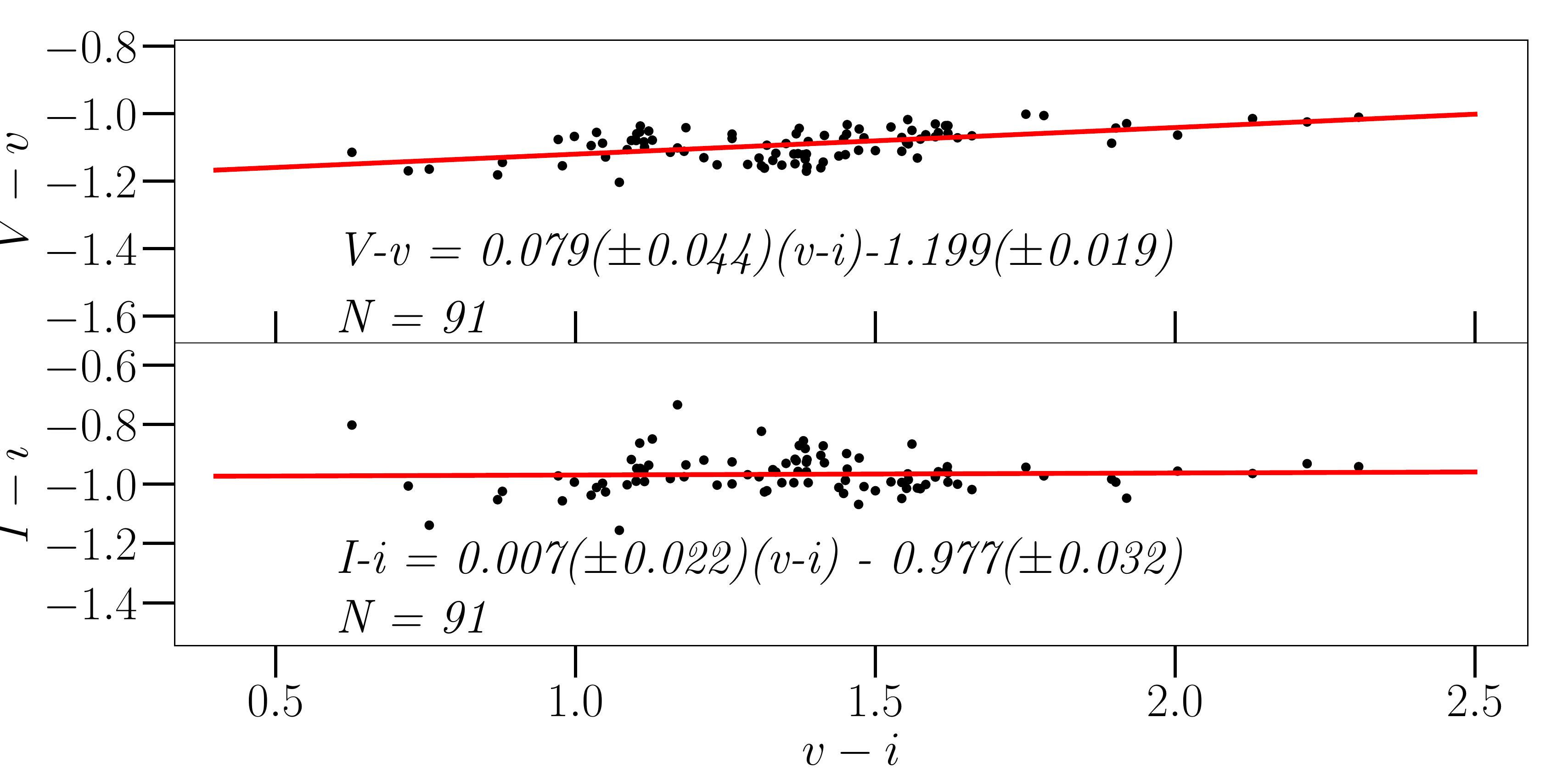}}
\caption{Transformation relations obtained for the $V$ and $I$ filters between the instrumental and the standard photometric systems. To carry out the transformation, we made use of a set of 94 and 91 standard stars for IAC (top panel) and Hanle (bottom panel) respectively, in the field of NGC 6712. The Hanle $I$-band color term is not significant, then a transformation of the form $I = 1.024 (\pm 0.006) i -1.354 (\pm0.093)$ was employed}

\label{transV}

\end{figure}

In order to transform our light curves from the $v$ instrumental magnitudes to the Johnson-Kron-Cousins standard \emph{V} system, we made use of the standard stars in the field of NGC 6712 which are included in the online collection of \citet{Stetson2000}\footnote{\url{http://www3.cadc-ccda.hia-iha.nrc-cnrc.gc.ca/community/STETSON/standards}} These stars are generally distributed in the cluster periphery. Unfortunately, this collection does not have the standard values for the $I$ filter.
Dr. Ra\'ul Michel from the Observatorio Astron\'omico Nacional, M\'exico, made available to us his bulky observations of stars in the field of NGC 6712, that have been transformed into the $I$-standard system using the equatorial standards of \citet{Landolt1983}. We used 94 of these stars as standards to transform our IAC CCD photometry. For the Hanle set of data we identified 91 standard stars. 
The mild colour dependence of the standard minus instrumental magnitudes is shown in Fig. \ref{transV} for our observations. The transformation equations in both \emph{VI} filters are explicitly given in the figure itself. We note that for the Hanle $I$-band transformation, the colour term is not significant and found that a linear transformation of the form $I = 1.024 (\pm 0.006) i -1.390 (\pm0.092)$ matches better the IAC and the Hanle $I$ light curves.

\section{Star membership using \textit{Gaia}}
\label{membership}

\begin{figure*}
\centerline{\includegraphics[width=18cm,height=9.cm]{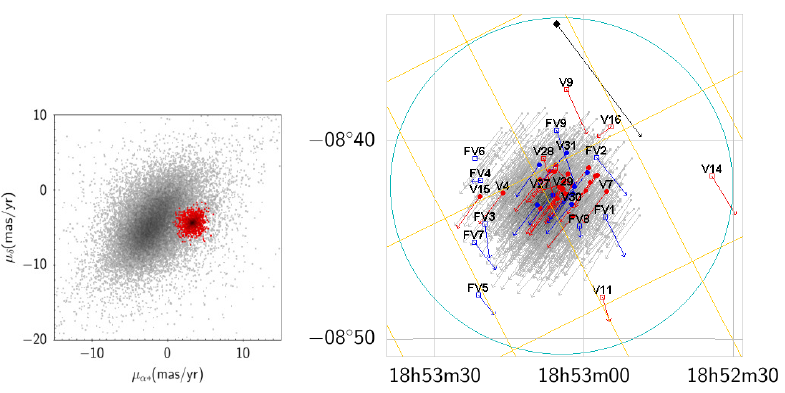}}
\caption{Left panel: Vector Point Diagram of the stars within a 8.5' radius. Grey dots are field stars and red dots are cluster members. Right panel: Projection on the sky of the proper motion vectors. Grey arrows are used for member stars, red arrows for known cluster variables, blue arrows for new variable stars reported in this work, red open circles for known field variables and blue open circles for new field variables. The yellow grid corresponds to the Galactic $l$ and $b$ coordinates. The large black arrow indicates the direction of the Galactic center. Vectors have been enlarged 20000x for visualisation purposes. See $\S$  \ref{dynamics} for further discussion.}
\label{vpd} 
\end{figure*}

To determine cluster membership of the stars in our FoV, we used the high-quality astrometric data available in \textit{Gaia}-DR2 \citep{Gaia2018}. We employed the procedure developed by \citet{Bustos2019}, consists of two stages: the first stage is based on the Balanced Iterative Reducing and Clustering using Hierarchies (BIRCH) algorithm \citep{Zhang1996} in a four-dimensional space of physical parameters -gnomonic projection of celestial coordinates and proper motions- that detects groups of stars in this 4D space; in the second stage an analysis of the projected spatial distribution of stars with different proper motions allows the extraction of most of the members in the outskirts of the cluster or with large proper motions dispersion. Finally, it is checked that the extracted cluster members have their positions in celestial coordinates, the Vector Point Diagram (VPD) of the proper motions and the CMD (of Gaia photometric system) consistent with the characteristics of a globular cluster. We used a tidal radius of 8.5 arcmin from \citet{Harris1996} (2010 edition) to set a boundary for the cluster and found 60 447 measured stars which we then cross-matched with our own data set. We found 1529 likely cluster members, for 1100 of which we possess light curves. In Fig. \ref{vpd} we display the Vector Point Diagram (VPD) resulting from this membership analysis. The axes in the VPD correspond to the components of the proper motion vectors, therefore every proper motion is represented by a point. Since the cluster members share a common motion, they appear as a small concentration different from the wider distribution of field stars.
Further comments on the cluster dynamics will be given later in $\S$ \ref{dynamics}.

\section{Variable stars in NGC 6712}
\label{variable_stars_6712}

\begin{table}
\scriptsize
\begin{center}
\caption{Time-series \textit{VI} photometry for the variables stars observed in this work$^*$}
\label{tab:vi_phot}
\centering
\begin{tabular}{cccccc}
\hline
Variable &Filter & HJD & $M_{\mbox{\scriptsize std}}$ &
$m_{\mbox{\scriptsize ins}}$
& $\sigma_{m}$ \\
Star ID  &    & (d) & (mag)     & (mag)   & (mag) \\
\hline
 V1 & $V$& 2455723.42031& 16.611 & 17.723 & 0.006 \\   
 V1 & $V$& 2455723.42540& 16.601 & 17.714 & 0.007 \\
\vdots   &  \vdots  & \vdots & \vdots & \vdots & \vdots  \\
 V1 & $I$ & 2455723.41738 & 15.622 & 16.608 & 0.008\\  
 V1 & $I$ & 2455723.42325 & 15.631 & 16.617 & 0.012  \\ 
\vdots   &  \vdots  & \vdots & \vdots & \vdots & \vdots  \\
 V2 & $V$ & 2457555.59991 & 12.610& 15.120 & 0.001 \\   
 V & $V$ & 2457555.60211 & 12.611&  15.121&  0.001 \\
\vdots   &  \vdots  & \vdots & \vdots & \vdots & \vdots  \\
 V2 & $I$ & 2457555.60101 & 10.564&  13.386 & 0.001 \\    
 V2 & $I$ & 2457555.60322 & 10.553&  13.375 & 0.001 \\   
\vdots   &  \vdots  & \vdots & \vdots & \vdots & \vdots  \\
\hline
\end{tabular}
\end{center}
* The standard and
instrumental magnitudes are listed in columns 4 and~5,
respectively, corresponding to the variable stars in column~1. Filter and epoch of
mid-exposure are listed in columns 2 and 3, respectively. The uncertainty on
$\mathrm{m}_\mathrm{ins}$ is listed in column~6, which also corresponds to the
uncertainty on $\mathrm{M}_\mathrm{std}$. A full version of this table is available at the CDS database.

\end{table}

\subsection{Search for new variables}
\label{search}

In order to identify the variable star population in NGC 6712, we used several methods that we describe below. Our first approach was the use of the string-length method (\citealt{Burke1970}, \citealt{Dworetsky1983}). We phased each light curve in our data with periods between 0.02 d and 1.7 d in steps of $10^{-6}$ d. In each case, the string-length parameter $S_{Q}$ was calculated. The best phasing occurs when $S_{Q}$ is minimum, and corresponds to the best period in our data. We then created a plot of the obtained minimum $S_{Q}$ vs X-coordinate in our reference image for each light curve in our collection (Fig. \ref{sq}). 
All variables in Table \ref{member_variables} and in Table \ref{field_variables} are identified. We note that most of the variables are located below an arbitrary threshold at 0.4, hence we individually explored each light curve below this value. Using this method, we identified two new cluster variables (V30 and V31) and seven new field variables (FV), FV1-FV7, all of which seem to be contact binaries or of the EW-type. 

The second method was to explore the regions of the CMD where variable stars are expected to be found. In particular, we explored the stars near the tip of the Red Giant Branch (TRGB) and we found 4 new semi-regular or SR variables (V32-V35), a SR-type field variable (FV8) and a SR candidate (which we will denote as C) for which we are unsure of its variation.

The last method used was to plot the rms vs. mean magnitude of all the stars in our FoV (Fig. \ref{rms}). We performed the search in regions where variable stars such as RR Lyrae and SR type stars are usually located. With this approach we identified two more SR-type variable stars (V36, and FV9). 

\begin{figure*}
\centerline{\includegraphics[width=17cm,height=21.cm]{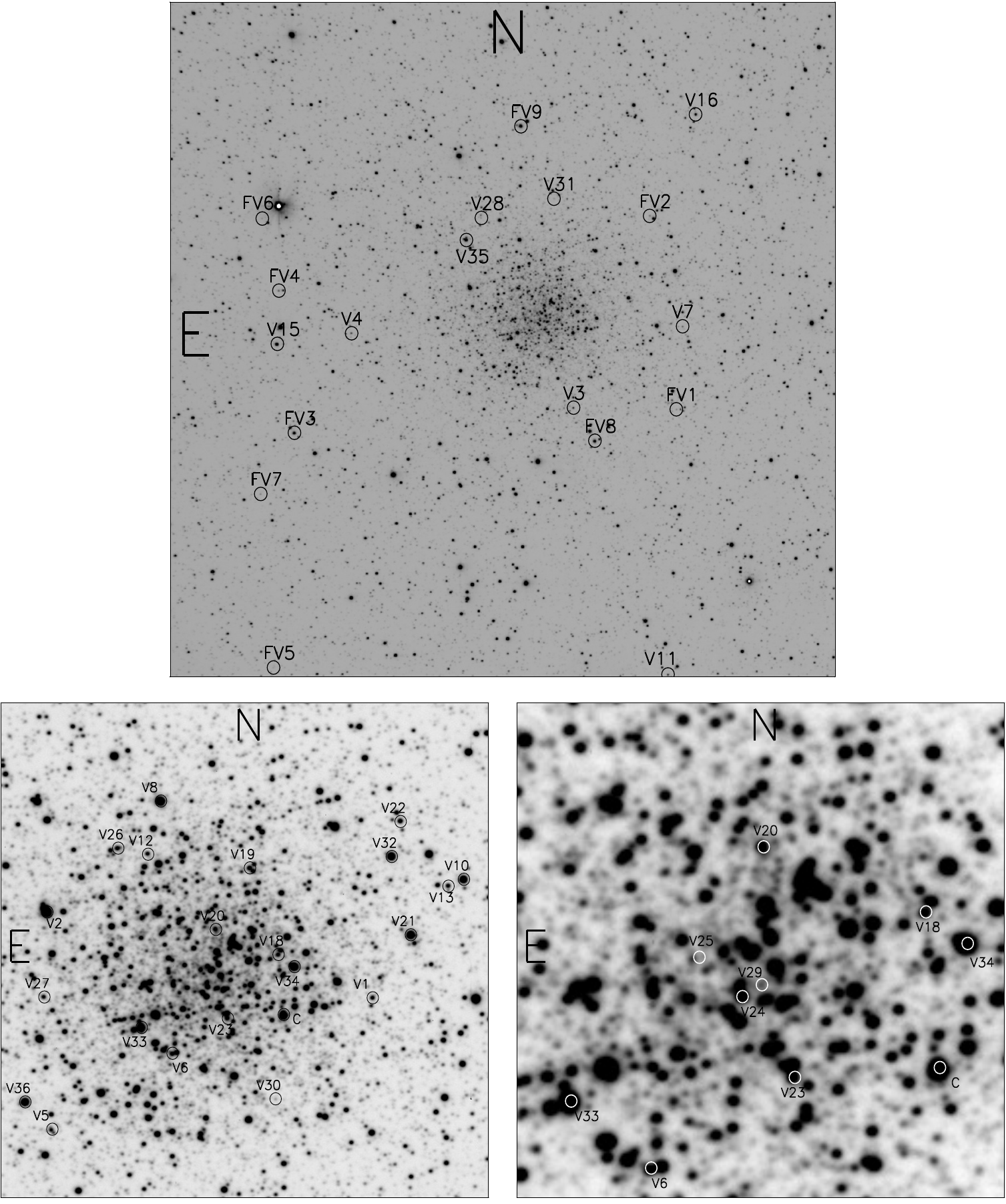}}
\caption{Identification chart for all known and newly discovered variables in NGC 6712. The bottom two diagrams show the central regions of the cluster and some stars are identified in more than one panel: The field sizes of the three panels are 10.4$\times$10.4, 3.3$\times$3.3 and 1.3$\times$1.3 arcmin$^2$, respectively. Variables labelled 'FV' are field stars according to our proper motion analysis (see $\S$ \ref{membership}). The marker V25 shows no star but this is an X-ray source (LMXB) \protect\citep{Homer1996} not visible in our images.}
\label{IDCHART} 
\end{figure*}

\begin{figure}
\centerline{\includegraphics[width=8.1cm, height=8.1cm]{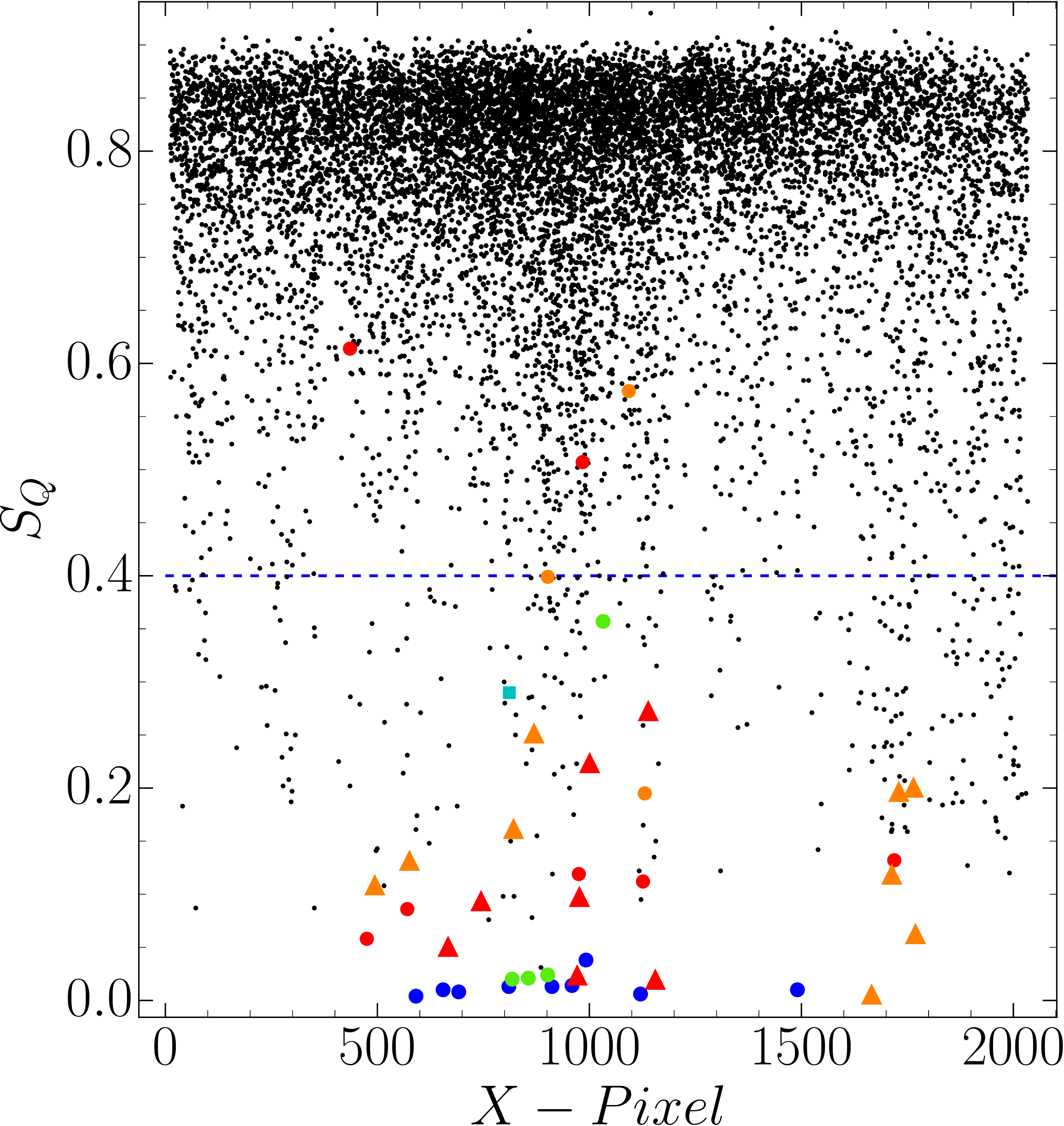}}
\caption{Minimum value for the string-length parameter $S_{Q}$ calculated for the 11,294 stars with a light curve in our $V$ reference image, versus CCD X-pixel coordinate. The blue circles correspond to a RRab stars, green circles to RRc stars, red circles and triangles correspond to semi-regular variables and orange circles and triangles to EW stars. The cyan square denotes a SR-variable candidate. Triangles are used for the newly discovered variables. The dashed blue line is an arbitrary threshold set at 0.4, below which most of the known variables are located. See  $\S$  \ref{search} for a discussion.}
\label{sq} 
\end{figure}

\begin{figure}

 \includegraphics[width=8.0cm,height=9.5cm]{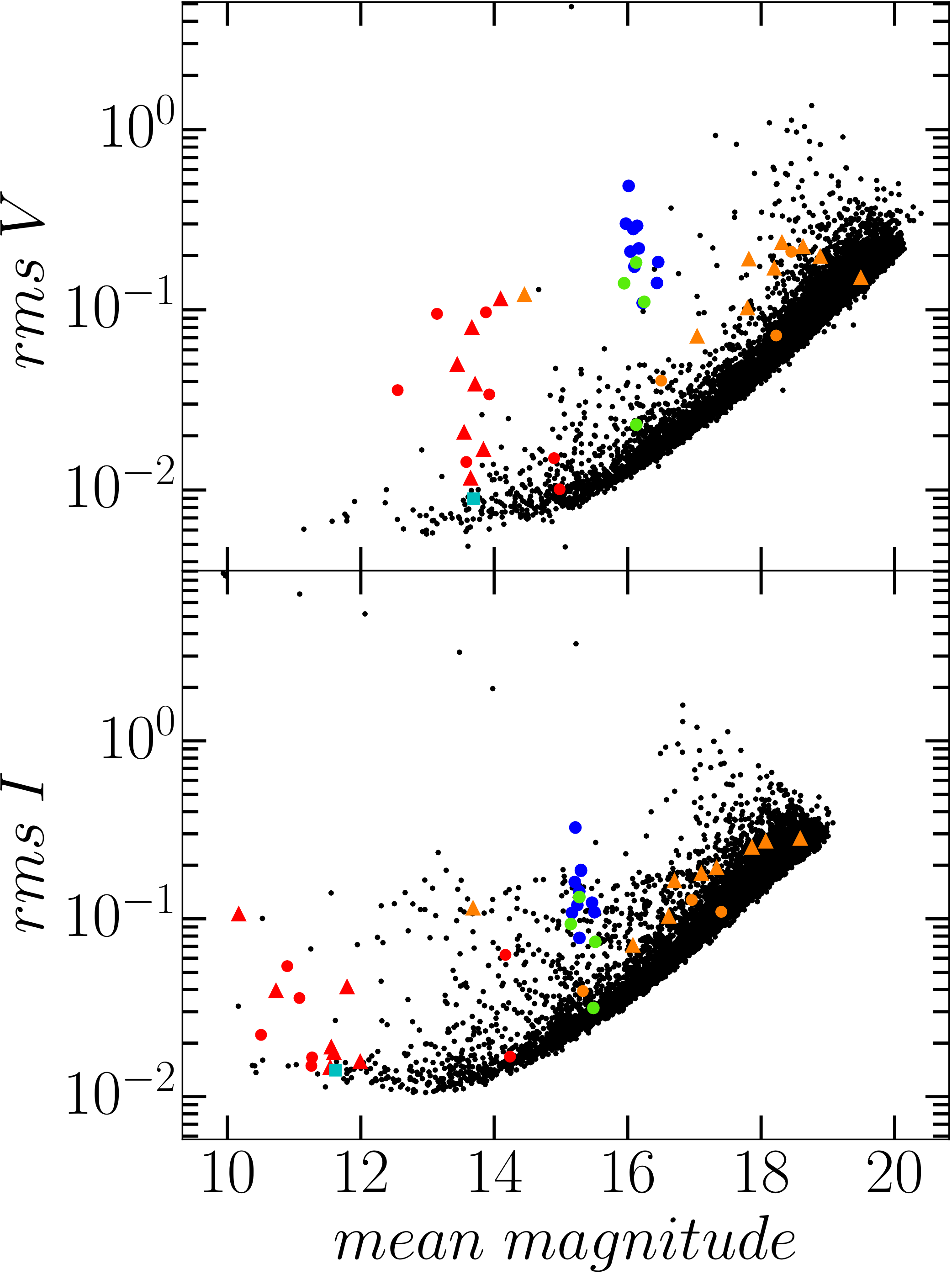}
\caption{The rms magnitude deviations as a function of the mean magnitudes $V$ and $I$.  The blue circles correspond to a RRab stars, green circles to RRc stars, red circles correspond to semi-regular variables and orange circles and triangles to EW stars. The cyan square denotes a SR-variable candidate. Triangles are used for the newly discovered variables.} 
\label{rms}

\end{figure}

In Fig. \ref{IDCHART} we present an identification chart with all known and newly found variables at present, including those that may not be cluster members. It is important to mention that in our data set we were not able to recover the light curves of V9 and V14 since they are out of our FoV, V11 since it lies near the lower border of our reference image and was not measured, and V25 
since it is below the limit of our photometry. Notice that
V25 is an X-ray source (LXMB), with a faint optical counterpart. We also found that several of the known variable stars are non-cluster members, namely, V9, V11, V14, V16, V17, V27, V28 and V29, according to the method used to determine stellar membership described in $\S$  \ref{membership}. 
The time-series \emph{VI} photometry
obtained in this work is reported in Table \ref{tab:vi_phot},
of which only a small portion is included in the
printed version of the paper. The full table shall be
available in electronic form in the Centre de Donn\'es
astronomiques de Strasbourg database (CDS).

\begin{table*}
%\scriptsize
\begin{center}
\caption{Data of member variable stars in NGC 6712 in the FoV of our images.}
\label{member_variables}

\begin{tabular}{llccccccccc}
\hline
Star ID & Type & $\big<V\big>$ & $\big<I\big>$   & $A_V$  & $A_I$ & $P$    &  $\rm HJD_{\rm max}$ & $\alpha$ (J2000.0)  & $\delta$  (J2000.0)   &  \textit{Gaia}-DR2 Source    \\
        &      & (mag) & (mag)   & (mag)  & (mag) & (days) &  + 2450000 &                     &         \\
\hline
V1  & RRab      & 16.31     & 15.40     & 1.11  & 0.75   & 0.512039   & 7574.6054     & 18:52:59.85 & -08:42:31.3 & 4203848946351463808 \\
V2  & SR        & 12.55     & 10.50     & >0.15 & >0.07  & 109.0$^1$  & --       & 18:53:08.79 & -08:41:56.8 & 4203849393028208384 \\
V3  & RRab      & 16.24     & 15.26     & 0.57  & 0.38   & 0.655956   & 7574.6678     & 18:53:02.29 & -08:43:47.4 & 4203848877631995136 \\
V4  & RRab      & 16.47     & 15.41     & 0.56  & 0.36   & 0.611745   & 7574.6268     & 18:53:16.29 & -08:42:38.1 & 4203849122517862144 \\
V5  & RRab      & 16.41     & 15.44     & 1.14  & 0.74   & 0.545362   & 7572.6357     & 18:53:08.68 & -08:43:24.1 & 4203848778920519168 \\
V6  & RRab \textit{Bl}     & 16.45     & 15.56     & 0.86  & 0.59   & 0.510871   & 7553.6906     & 18:53:05.35 & -08:42:53.6 & 4203848985005261312 \\
V7  & M         & 16.98$^7$ & 11.97$^7$ & >0.62 & >0.23  & 193.0$^2$  & --        & 18:52:55.39 & -08:42:32.6 & 4203849672273972864 \\
V8  & SR        & 13.17$^7$ & 10.90$^7$ & >0.48 & >0.28  & 116.29$^2$ & --       & 18:53:05.66 & -08:41:12.4 & 4203849461747677056 \\
V10 & L         & 13.94$^7$ & 11.27$^7$ & >0.21 & --  &   --     & --       & 18:52:57.35 & -08:41:43.9 & 4203849874064515456 \\
V12 & RRab      & 16.30     & --        & 1.21  & --   & 0.502790   & 5284.7907     & 18:53:06.02 & -08:41:33.4 & 4203849088158338944 \\
V13 & RRab      & 16.13     & 15.26     & 1.02  & 0.66   & 0.562655   & 7574.6637     & 18:52:57.77 & -08:41:46.5 & 4203849809712937088 \\
V15 & L         & 13.83$^7$ & 11.08$^7$ & >1.5 & --  &  --     & -             & 18:53:20.99 & -08:42:48.2 & 4203837401469330944 \\
V18 & RRc       & 16.15     &   15.32   & 0.47  & 0.28 & 0.353543     & 7569.5876     & 18:53:02.44 & -08:42:14.0 & 4203849053798612096 \\
V19 & RRc       & 16.00     & 15.19     & 0.40  & 0.24   & 0.412161   & 7586.6633     & 18:53:03.22 & -08:41:39.2 & 4203849844072610816 \\
V20 & RRc  & 16.20  & 15.48 & 0.34  & 0.25    & 0.250521 & 5779.3246 & 18:53:04.16 & -08:42:03.7 & 4203849083790434304 \\
V21 & L         & 13.58$^7$ & 11.26$^7$ & >0.33 & --  &  --     & --  & 18:52:58.80 & -08:42:06.2 & 4203849805344920576 \\
V22 & RRab$^{3}$ & 16.17     & 15.24     & 0.58  & 0.41  & 0.654789   & 7586.6187   & 18:52:59.08 & -08:41:20.5 & 4203849874064518016 \\
V23 & RRab$^{4}$ & 16.23     & 15.26     & 0.31 & 0.19   & 0.642451   & 7572.6241     & 18:53:03.83 & -08:42:39.4 & 4203848985005549056 \\
V24 & RRab       & 16.23     & 15.25     & 0.99 & 0.61   & 0.576454   & 7586.6334     & 18:53:04.37 & -08:42:26.9 & 4203849053725057280 \\
V26 & RRc       & 16.16     & 15.64      & 0.12  & 0.09   & 0.334317   & 7569.6551     & 18:53:06.80 & -08:41:31.2 & 4203849466042002944 \\
V28 & EW        & 18.23     & 17.42     & 0.26 & 0.08    & 0.434497   & 7573.6734$^7$ & 18:53:08.09 & -08:40:52.2 & 4203849466042155264 \\
V29 & EW        & 18.45     & 16.95     & -    & -       & 0.453571   & 7569.6226$^7$ & 18:53:04.17 & -08:42:25.0 & 4203849053798521472 \\
V30$^{5}$ & EW  & 18.68     & 17.91     & 0.61 & 0.49    & 0.504748   & 7567.6482$^7$ & 18:53:02.51 & -08:43:11.9 & 4203848881926249600 \\
V31$^{5}$ & EW  & 18.53     & 17.74     & 0.40 & 0.38    & 0.418762   & 7573.6734$^7$ & 18:53:03.49 & -08:40:34.4 & 4203850221956827648 \\
V32$^{5}$ & SRs  & 13.72$^6$ & 11.56$^6$ & >0.14 & >0.06  & 31.1   & 7586.6252  & 18:52:59.32 & -08:41:34.7 & 4203849805345042176 \\
V33$^{5}$ & SR  & 13.68$^6$ & 11.57$^6$ & >0.24 & --  & --          & --             & 18:53:06.20 & -08:42:43.2 & 4203848985079013760 \\
V34$^{5}$ & SR  & 13.54$^6$ & 11.59$^6$ & >0.25 & --  & --     & --       & 18:53:01.99 & -08:42:18.8 & 4203849049430686336 \\
V35$^{5}$ & SR  & 13.93$^6$ & 12.00$^6$ & >0.17 & --  & --        & --  & 18:53:09.02 & -08:41:12.4 & 4203849461747685760 \\
V36$^{5}$ & SRs  & 13.66$^6$ & 11.79$^6$ & >0.22 & >0.11  & 17.8 & 5724.4243 & 18:53:09.39 & -08:43:12.9 & 4203848808912547840 \\ %V44
C$^{8}$   & SRs & 13.84$^6$ & 11.64$^6$ & >0.21 & --  & 18.9& 5780.3438          &  18:53:02.30 & -08:42:38.4 & 4203849049430685312 \\ %V40
\hline

\hline
\end{tabular}
\raggedright
\center{
1. The period is from \citet{Oosterhoff1943}. 2. The period is from \citet{Sloan2010}. 3. Classified as RRab in this work. See $\S$  \ref{V22}. 4. Reclassified as RRab in
this work. See $\S$ \ref{v23}. 5. New variable discovered in this work.
6. Magnitude-weighted mean. 7. Time of primary minimum light. 8. Member variable candidate. \textit{Bl} denotes Blazhko effect.}

\end{center}
\end{table*}

\begin{table*}
%\scriptsize
\begin{center}
\caption{Data of known, unclassified and newly discovered field variable stars (FV) in NGC 6712 in the FoV of our images.}
\label{field_variables}

\begin{tabular}{llccccccccc}
\hline
Star ID & Type & $\big<V\big>$ & $\big<I\big>$   & $A_V$  & $A_I$ & $P$    &  $\rm HJD_{\rm min}$ & $\alpha$ (J2000.0)  & $\delta$  (J2000.0)   &  \textit{Gaia}-DR2 Source    \\
        &      & (mag) & (mag)   & (mag)  & (mag) & (days) &  + 2450000 &                     &         \\
\hline

V16 & ?         & 14.98$^7$ & 14.16$^7$ & >0.06 & >0.06  &      --    & --   & 18:52:54.55 & -08:39:17.0 & 4203850187670201088 \\ 
V17 &  ?        & 14.89$^7$ & 14.23$^7$ & >0.24 &--  & --       & --  & 18:53:05.86 & -08:41:23.4 & 4203849083790555520 \\ 
V27 & EW$^{1}$  & 16.51     & 15.34     & 0.12 & 0.08    & 0.425714   &       7586.6325$^2$ & 18:53:08.88 & -08:42:30.7 & 4203849019364947456 \\
FV1$^{4}$  & EW  & 18.25     & 17.14     & 0.54 & 0.48    & 0.354499   & 7574.6363$^2$ & 18:52:55.78 & -08:43:49.1 & 4203847438817518976 \\ %V30
FV2$^{4}$  & EW  & 17.84     & 16.65     & 0.27 & 0.25    & 0.355390   & 7567.5873$^2$ & 18:52:57.45 & -08:40:50.4 & 4203849874064515840 \\ %V31
FV3$^{4}$  & EW  & 14.44     & 13.66     & 0.38 & 0.35    & 0.445495   & 7574.6047$^2$ & 18:53:19.90 & -08:44:10.1 & 4203837298410747776 \\ %V34
FV4$^{4}$  & EW  & 17.04     & 16.07     & 0.23 & 0.19    & 0.368598   & 7572.6428$^2$ & 18:53:20.86 & -08:41:58.9 & 4203849259956801792 \\ %V35
FV5$^{4}$  & EW  & 18.93     & 17.96     & 0.54 & 0.54    & 0.347863   & 7572.6380$^2$ & 18:53:21.25 & -08:47:46.5 & 4203835756498276608 \\ %V36
FV6$^{4}$  & EW  & 18.39     & 17.41     & 0.40 & 0.39    & 0.462812   & 7573.8535$^2$ & 18:53:21.90 & -08:40:52.2 & 4203849362961930624 \\ %V37
FV7$^{4}$  & EW  & 17.88     & 16.74     & 0.72 & 0.51    & 0.344416   & 7567.6514$^2$ & 18:53:22.04 & -08:45:06.3 & 4203836718651092352 \\ %V38
FV8$^{4}$  & SRs  & 14.17$^3$ & 10.74$^3$ & >0.58 & --  & 12.3   &  7555.6001    & 18:53:00.93 & -08:44:18.1 & 4203848843272252160 \\ %V42
FV9$^{4}$  & SRs  & 13.45$^3$ & 10.17$^3$ & >0.15 & --  &24.6  & 8217.4473         & 18:53:05.58 & -08:39:27.6 & 4203850325109041792 \\ %V43
 \hline

\hline
\end{tabular}
\raggedright
\center{
1. Reclassified as EW in this work. See $\S$  \ref{V27}. 2. Time of primary minimum light. 3. Magnitude-weighted mean. 4. New variable discovered in this work. }

\end{center}
\end{table*}

\subsection{RR Lyrae stars in NGC 6712}
\label{RRLyr}

\begin{figure*} 
\centerline{\includegraphics[width=23cm, height=20cm]{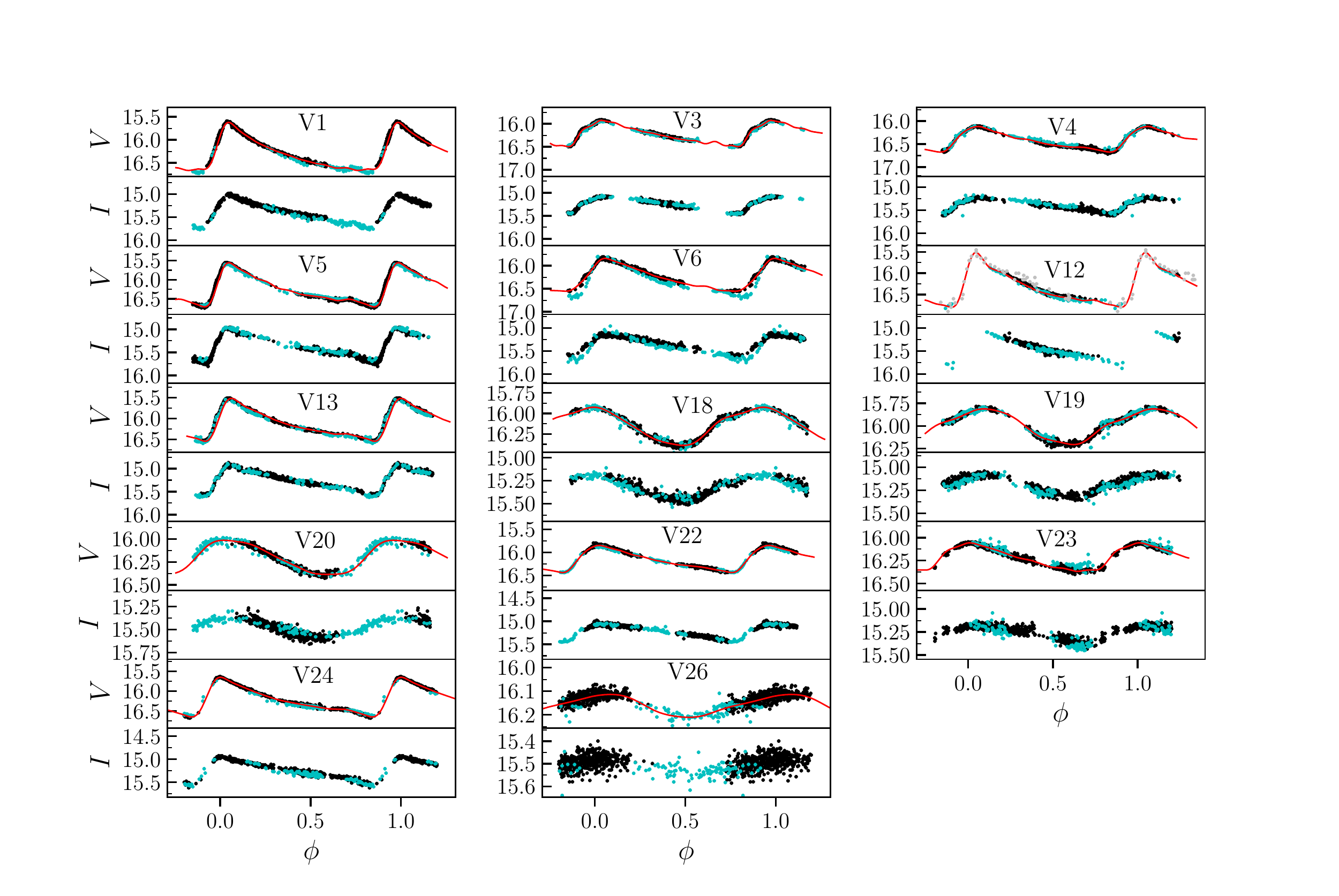}}
\caption{RR Lyrae stars in NGC 6712. Black and cyan symbols correspond to observations from IAC and Hanle respectively. The red line represents the Fourier fit. Note that the scale in the Y-axis is different for RRab and RRc stars. The grey dots in V12 come from \protect\citet{Sandage1966} and were used to complete our light curve in the $V$-band. (See $\S$ \ref{decomposition}).}
\label{grid_rr}
\end{figure*}

In the 2015 edition of the CVSGC, there are listed 8 RRab, 6 RRc and one RR Lyrae without mode classification. The present data and a careful inspection of the resulting light curves, made necessary it to reclassify some stars, namely V22, V23 and V27, as RRab, RRab and EW, respectively. The details of these new classifications are explained in  Appendix A. As a result, the RR Lyrae population of NGC 6712 is formed by 10 RRab and 4 RRc stars.
Our observations generally cover the light curves phase reasonably well, with the exception of V12,
the reason being its nearly half-day periodicity that operates against a better phase coverage. In order to confirm its light curve shape and pulsation mode type, we complemented our $V$ data with $V$ data of  \citet{Sandage1966}. While the scatter of these observations is large, they confirm the RRab-type nature of this star. In spite of this, V12 shall not be considered for the physical parameters calculations.
With the inclusion of the data from Hanle, we were also able to detect a mild modulation in the amplitude of the light curve of V6. This suggests the presence of a previously undetected Blazhko effect.
The light curves of all RR Lyrae are shown in Fig. \ref{grid_rr}. 

\subsubsection{Bailey diagram and Oosterhoff type}
\label{baileyDiagram}

\begin{figure}
\begin{center}
 \includegraphics[width=7.5cm,height=12cm]{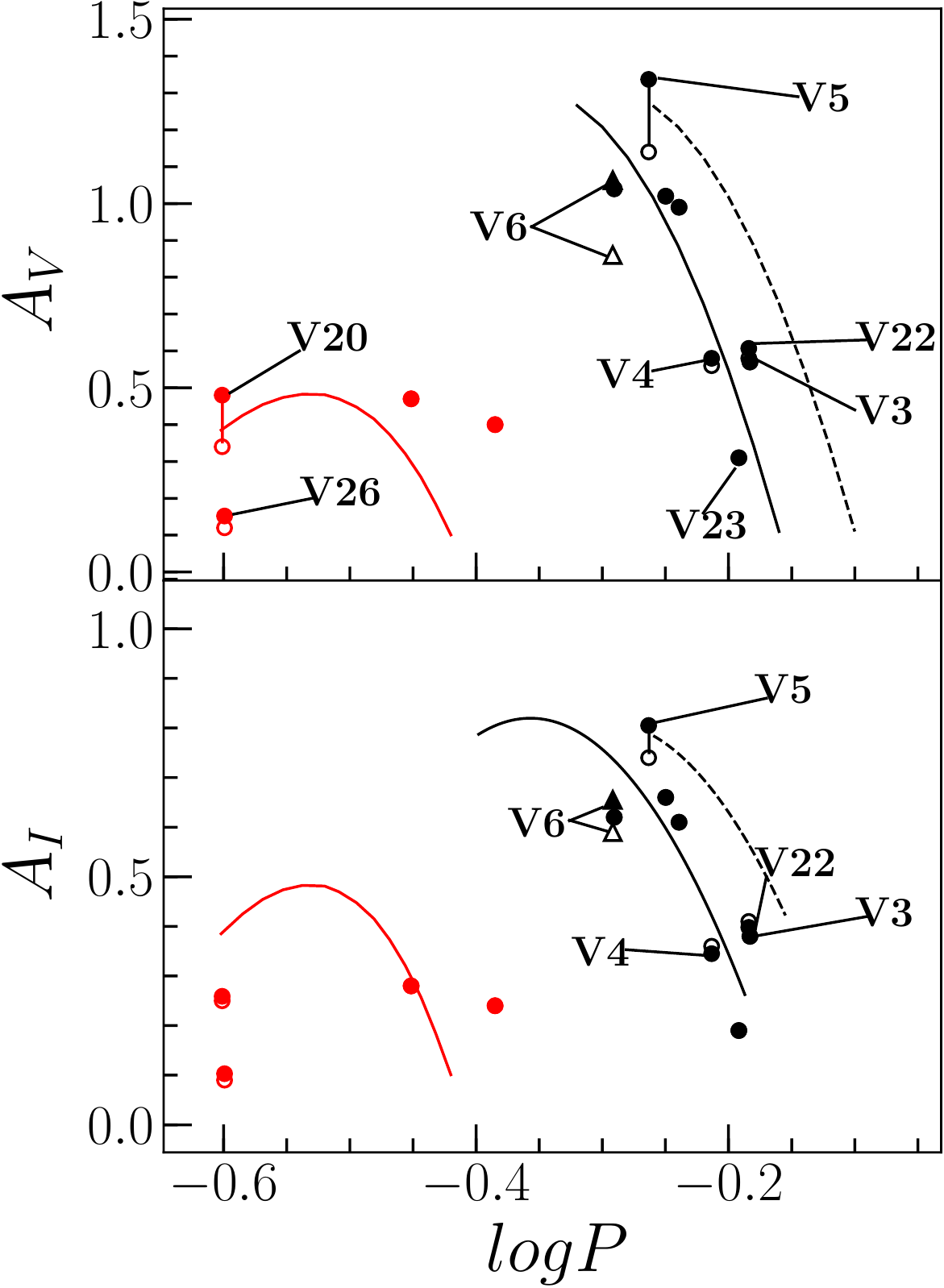}
\caption{Bailey diagram for NGC 6712. Filled black and red symbols represent RRab and RRc stars respectively. The open symbols denote the original position of the RR Lyrae before a correction to their amplitude was performed due to contamination of a neighbour star. The triangular marker (V6) is a Blazhko RRab star. The continuous and dashed black lines in the top panel are the loci for unevolved and evolved stars according to \protect\citet{Cacciari2005}. The red parabolas were calculated by \protect\citet{Arellano2015} from RRc stars in five OoI clusters. In the bottom panel, the black solid and segmented loci for unevolved and evolved stars respectively are from \protect\citet{Kunder2013a}. See $\S$  \ref{baileyDiagram} for a more detailed discussion.} 
\label{bailey}
\end{center}

\end{figure}

The period-amplitude plane for RR Lyrae stars, also known as the Bailey diagram, is a useful tool as it clearly segregates pulsation modes,
helps defining the Oosterhoff type of a given cluster, and identifies RRab stars that may be advanced in their evolution towards the asymptotic giant branch (AGB). The diagram for NGC 6712 is shown in Fig. \ref{bailey} for the \emph{VI} band passes. The periods and amplitudes are listed in Table \ref{member_variables}. The amplitudes were measured from the corresponding fit provided by the Fourier decomposition of their light curves (See $\S$  \ref{decomposition}). The average of the periods of the RRab stars in NGC 6712 is $\big <P_{ab} \big >$ = 0.58 $\pm$ 0.02, which classifies it as a OoI-type cluster.
 The RRab and RRc stars in this diagram occupy the loci that has been observed to be consistent with the definition of a OoI-type globular cluster (black solid lines in $V$ and $I$ filters, respectively). V6 is classified as a RRab-type star, nevertheless, it is located at an odd position on the Bailey diagram, which can be explained due to its Blazhko nature. We note that V5 is located along the evolved star sequence, however, its position of the CMD lies very close to the ZAHB and without a formal analysis of a possible secular period change, we cannot state its evolutionary status relative to the rest of the RR Lyrae stars.\\

\subsection{Semi-regular variable stars in NGC 6712}
\label{sr_stars}

\begin{figure*} 
\includegraphics[width=16cm, height=17.0cm]{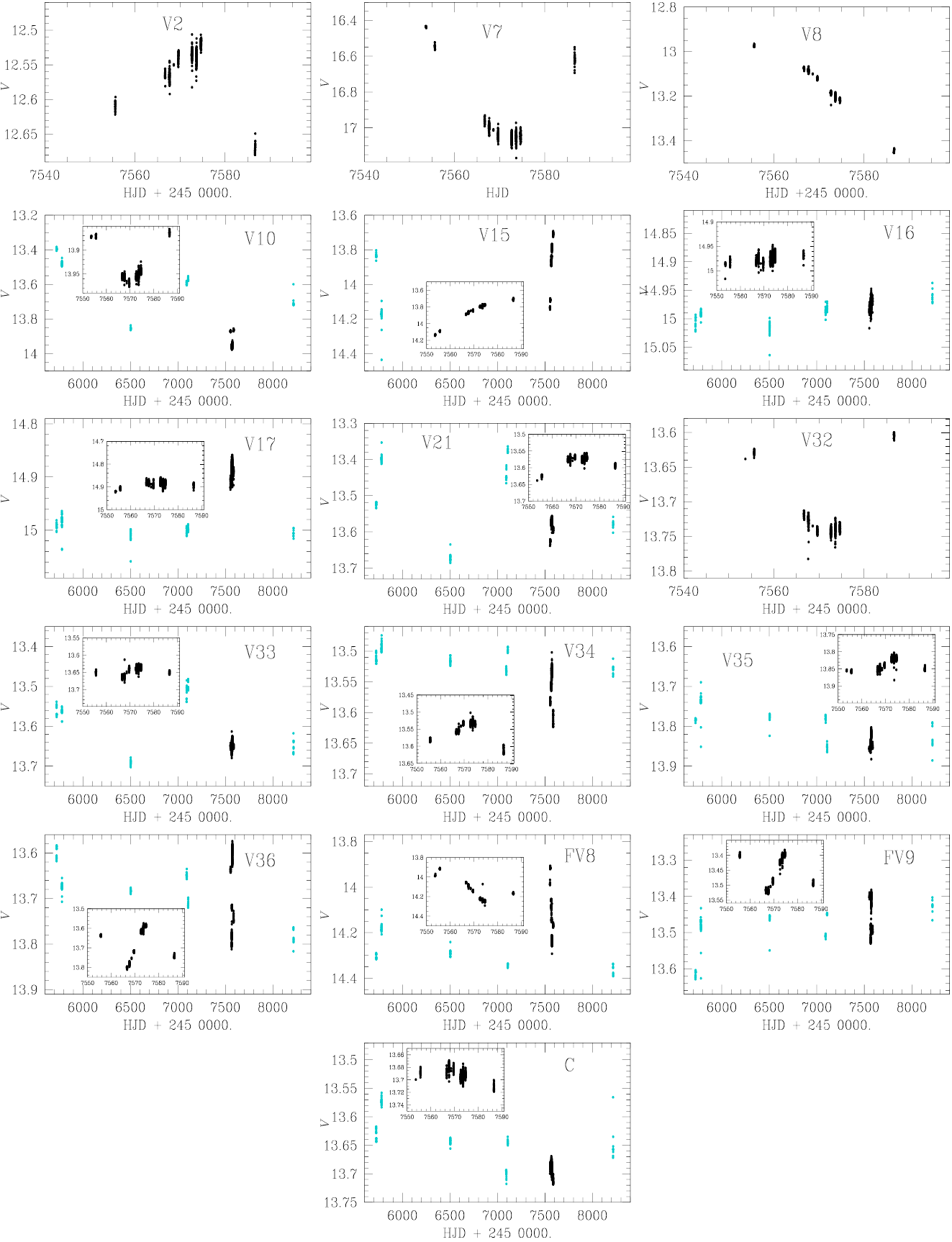}
\caption{SR stars in NGC 6712. Colour symbols are as in Fig. \ref{grid_rr}. The full time-span is of nine years and the long-term variation is evident. The insets display a blow up of the IAC data (black symbols) for which a smoother and continuous pace is available, and the variation in a shorter time span is shown. Stars with no inset lack Hanle data due to saturation. See $\S$ \ref{sr_stars}.}
    \label{high-amp}
\end{figure*}

There are seven known semi-regular or long-period variables, classified as SR or L in the CVSGC, namely, V2, V7, V8, V10, V14, V15 and V21. Unfortunately V14 lies outside our FoV and was not measured. Without a sufficiently long time-base and dense data set, the classification of long-term semi-regular variables is always dubious. Our Hanle and IAC data sets span eight years, which enables to peer the long-term light behaviour. On the other hand, the more continuous pace of the observations at the IAC, enabled us to analyse even the short-term behaviour. 
Additionally, we found seven semi-regular variables and have tagged them as SR-type stars. Five of these stars are cluster members (V32-V36) and two are not (FV8 and FV9).
In Fig. \ref{high-amp}  the light curves of the above stars are displayed. The insets in the figure show the short-term light curve behaviour and sometimes suggest a periodicity. A proper period search in all these light curves allowed us to find reliable periods in five of these stars, listed in Tables \ref{member_variables} and \ref{field_variables}. The corresponding phased light curves are displayed in Fig. \ref{SR_FASES}

\subsection{Eclipsing binaries in the field of NGC 6712}
\label{ew_stars}
\begin{figure*} 
\centerline{\includegraphics[width=20cm, height=18cm]{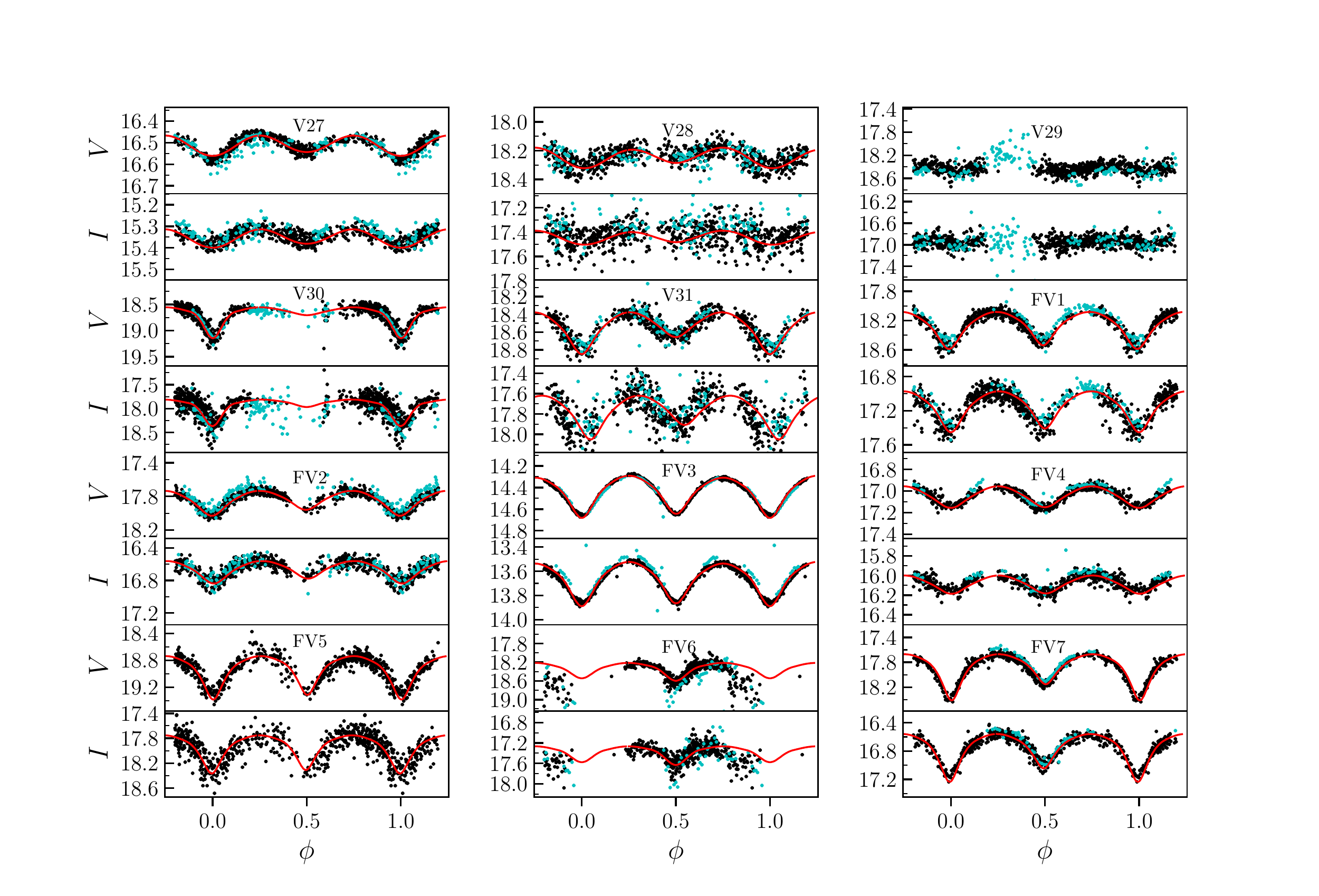}}
\caption{EW stars in NGC 6712. Black and cyan symbols correspond to observations from IAC and Hanle respectively. Red continuous curves are the model solution. The stars V30, V31 and FV1-FV7 are newly discovered variables. See $\S$ \ref{ew_stars}.}
    \label{ew}
\end{figure*}

The catalogue of \citet{Clement2001} lists two W Ursae Majoris-type binaries or EW stars (V28 and V29), first reported by \citet{Pietrukowicz2004}. We were able to recover their light curves from our data and to identify nine new eclipsing binaries present in our FoV. The $V$ and $I$ light curves are shown in Fig. \ref{ew}. The light curves present a morphology typical of the W UMa contact systems, except V30 which presents a steeper descent to minimum and is probably not a contact binary. V27 was previously misidentified as a RRc star (see $\S$  \ref{V27}). 

From the study of cluster membership using \textit{Gaia}-DR2, most
 of the new binaries seem to be field stars, and so their reddening and distances are unknown. In order to get some information from their photometric light curves we have assumed they are contact systems. Based on \textit{Gaia}-DR1, 
 \citet{Chen2018} derived period-luminosity relations for W UMa-type systems,
 of the form ~ $M_\lambda(max)_C = a_\lambda ~log~P ~+~ b_\lambda$, claiming 8\% distance accuracy, where $P$ is the orbital period. We have used the 3D Dust Mapping \footnote{http://argonaut.skymaps.info} facility 
 to get the increase of $E(B-V)$ with distance, looking for the best agreement between ~$M_\lambda(max)_C$~
 and ~$M_\lambda = m_\lambda(max) - A_\lambda - 5~log~d(pc) + 5$, for each one of the systems in the \emph{VI} bands.
 The $E(B-V)$ in the  direction of the cluster changes from $\sim$0.13 at 2 kpc, to $\sim$ 0.37-0.40 at 7.9 kpc.
 The results are given in the columns 2 and 3 of Table \ref{table_ew}, and are
 consistent with the \textit{Gaia}-DR2 membership study, supporting the hypothesis that most of them are closer than the cluster.
 
 With a value for $E(B-V)$, we can derive the intrinsic colours $(V-I)_o$ of the binaries, used
 to estimate the effective temperatures of the components from the calibration
 $(V-I)_o - T_{\rm eff}$ of \citet{Huang2015}, and the relative contribution of
 each component to the total flux in $V$ and $I$, calculated in the light curves fits.
  The \emph{VI} light curves have been modelled with the code BinaRoche \citep{Lazaro2009}, weighting 
 in the fits the deviations of both components from the mass-radius relations of \citet{Awadalla2005} for W UMa binaries.
The derived parameters of the fits are given in Table \ref{table_ew}, while the
observed and model light curves are shown in Fig. \ref{ew}. 
 The collected data of V30 cover only one of the eclipses, which has been
 arbitrarily considered to be the primary, with a steeper descent than other light curves,
 and could be a detached or semidetached binary.
 
 Given the assumptions implicit in the analysis, and the quality of the light curves,
 we prefer not to put an error bar in the tabulated values.
 A realistic estimation of the uncertainties can be about
 $\pm$ 300 ~K in the $T_{\rm eff,1}$ effective temperatures, affected by the $E(B-V)$ used
 to deredden the observed $(V-I)$ colour. The ratios $T_{\rm 2/1}= T_{\rm eff,2}/T_{\rm eff,1}$ and 
 $q= M_2/M_1$ are determined by the photometric quality of the light curves, and its uncertainty 
 can be about $\pm 0.02$. The most uncertain parameter is the primary mass ~$M_1$, but with these values the models put the systems at distances similar to those given in column 3.

\begin{table*}
%\scriptsize
\begin{center}
\caption{Physical Parameters of EW stars in NGC 6712 in the FoV of our images ($\S$  \ref{ew_stars}). Stars labelled FV are not cluster members.}
\label{table_ew}

\begin{tabular}{cccccccccc}

\hline
Star ID &  $E(B-V)$ & $d$~(kpc.)& $M_1$($M_\odot$) & $q$ &  $T_{\rm eff,1}$ (K) & $T_{2/1}$ & $R_{1}$ ($R_\odot$) &$R_{2}$ ($R_\odot$) & $i(^o)$ \\
\hline
 V27 & 0.21  &  2.8 & 1.37 & 0.93  & 5530& 0.94 & 1.26 & 1.12 & 39.6 \\  
 V28 & 0.36  &6.3 &1.30 & 0.40  &  7480 & 0.98 &  1.33 & 0.83 & 45.0  \\
 V30 & 0.37  & 7.9 & 0.48 & 0.60  & 9750 & 0.60 &  0.88 & 0.74 &  77.5  \\
 V31 & 0.37  &  6.3 &  0.45 & 1.0  & 9700 & 0.72 &  0.86 & 0.86 &  68.3  \\ 
 FV1 &  0.35  &  4.0 &  0.92 & 0.99  &  6200 & 0.99 &  1.00 & 1.00 &  74.8  \\  
 FV2 & 0.29  &  4.0 &  1.19 & 0.44  &  5840 & 0.92 &  1.16 & 0.80 & 63.6  \\ 
 FV3 &  0.15  &  1.2 &  1.24 & 0.98  & 6220 & 0.98 & 1.26 & 1.25 &68.0  \\
 FV4 & 0.19  & 3.2 & 0.95 & 0.87  &  5850 & 0.98 & 1.01 & 0.96 &  56.1  \\    
 FV5 & 0.41  & 7.0 & 0.77 & 0.99  &  7110 & 0.98 & 0.90 & 0.91 &  80.0  \\   
 FV6 & 0.36  &  6.3 & 1.40 & 0.85  &  7300 & 0.75 &  1.29 & 1.13 &  83.4  \\  
 FV7 & 0.21  &  3.2 &  0.80 & 1.0   &  5800 & 0.90 &  0.92 & 0.92 &  79.0  \\ 
\hline   
\end{tabular}
\end{center}
\end{table*}

\section{RR Lyrae stars: [Fe/H] and $M_V$ from light curve Fourier decomposition}
\label{decomposition}

The light curves of RR Lyrae stars exhibit periodic oscillations and are hence well represented by a Fourier series of the following form: 

\begin{equation}
    m(t) = A_{0} + \sum_{k=1}^{N}A_{k}cos(\frac{2\pi}{P}k(t-E_{0}) + \phi_{k})
	\label{eq:quadratic}
\end{equation}

\noindent
where $m(t)$ is the magnitude at time $t$, $P$ is the period of pulsation, and $E_0$ is the epoch.  To estimate the Fourier parameters we followed a least-squares approach, where the quantities to be minimised are the amplitudes $A_k$ and phases $\phi_k$ of the light curve components. The Fourier parameters, i.e., the amplitudes and phases of the harmonics in Eq. \ref{eq:quadratic} are defined as $R_{ij} = A_{i}/A_{j}$ and $\phi_{ij} = j\phi_{i} - i\phi_{j}$, respectively. The estimated Fourier coefficients for the RR Lyrae stars in our data are listed in Table \ref{tab:fourier_coeffs}.

\defcitealias{Kovacs2001}{KW01}
\defcitealias{Morgan2007}{M07}
\defcitealias{Jurcsik1996}{JK96}
\defcitealias{Nemec2013}{N13}

We have employed the calibrations of \citet{Jurcsik1996} (\citetalias{Jurcsik1996}) and \citet{Kovacs2001} to calculate the metallicity and absolute magnitude of the RRab stars. The specific set of equations used in this work is listed below. \\

\begin{equation} 
{\rm [Fe/H]}_{J} = -5.038 ~-~ 5.394~P ~+~ 1.345~\phi^{(s)}_{31},
\label{eq:JK96}
\end{equation} 

\begin{equation} 
M_V = ~-1.876~\log~P ~-1.158~A_1 ~+0.821~A_3 + K.
\label{eq:ḰW01}
\end{equation} 

\noindent  
These calibrations have an associated uncertainty of 0.14 dex and 0.04mag for the metallicity and the absolute magnitude respectively. Since the the metallicity is given in the Jurcsik-Kov\'acs scale, we can transform it to the standard Zinn-West scale \citep{Zinn1984} with the following equation: [Fe/H]$_{J}$ = 1.431[Fe/H]$_{ZW}$ + 0.88 \citep{Jurcsik1995}. For eq. \ref{eq:ḰW01}, we adopted the zero point $K$ = 0.41 from the analysis of \citet{Arellano2010}. \\

For the RRc stars, we adopted the calibrations given by  \citet{Morgan2007} and \citet{Kovacs1998}, respectively:

%\noindent

$${\rm [Fe/H]}_{ZW} = 52.466~P^2 ~-~ 30.075~P ~+~ 0.131~\phi^{2(c)}_{31}$$
\begin{equation}
~~~~~~~	~-~ 0.982 ~ \phi^{(c)}_{31} ~-~ 4.198~\phi^{(c)}_{31}~P ~+~ 2.424,
\label{eq:Morgan07}
\end{equation}

\begin{equation}
M_V = 1.061 ~-~ 0.961~P ~-~ 0.044~\phi^{(s)}_{21} ~-~ 4.447~A_4.
\label{eq:K98}	
\end{equation}

\noindent 
These calibrations have an associated uncertainty of 0.14 dex and 0.042mag for the metallicity and the absolute magnitude respectively.
One can transform the coefficients from cosine series phases into sine series using the following relation, if needed: 
\begin{equation}
\phi^{(s)}_{jk} = \phi^{(c)}_{jk} - (j - k) \frac{\pi}{2}.
\end{equation}

\begin{table*}
\begin{center}
\caption{Fourier coefficients $A_{k}$ for $k=0,1,2,3,4$, and phases from the cosine series $\phi^{(c)}_{21}$, $\phi^{(c)}_{31}$ and $\phi^{(c)}_{41}$, for RRab and RRc stars. The numbers in parentheses indicate the uncertainty on the last decimal place. Also listed is the deviation parameter $D_{\mbox{\scriptsize m}}$ for the RRab stars (see  \protect\citetalias{Jurcsik1996}).}
\label{tab:fourier_coeffs}
\centering                   
\begin{tabular}{cllllllllc}
\hline
Star ID     & $A_{0}$    & $A_{1}$   & $A_{2}$   & $A_{3}$   & $A_{4}$   &
$\phi^{(c)}_{21}$ & $\phi^{(c)}_{31}$ & $\phi^{(c)}_{41}$ & $D_{\mbox{\scriptsize m}}$ \\
     & ($V$ mag)  & ($V$ mag)  &  ($V$ mag) & ($V$ mag)& ($V$ mag) & & & & \\
\hline
       &       &   &   & RRab stars  &     & &           &                 &       \\
\hline
V1    & 16.311(2)  & 0.381(3) & 0.182(3) & 0.111(3) & 0.071(3) & 4.120(25) & 8.286(37)  & 6.185(55)   & 1.4      \\
V3    & 16.238(12) & 0.221(2) & 0.092(2) & 0.050(2) & 0.022(3) & 4.305(36) & 8.790(57)  & 7.336(114)  & 2.1      \\
V4    & 16.428(1)  & 0.209(2) & 0.097(2) & 0.056(2) & 0.018(2) & 4.211(27) & 8.660(49)  & 7.197(126)  & 5.7      \\
V5    & 16.253(2)  & 0.379(3) & 0.202(3) & 0.128(3) & 0.093(2) & 4.037(20) & 8.349(30)  & 6.482(58)   & 2.5      \\
V6    & 16.257(2)  & 0.292(2) & 0.118(2) & 0.052(3) & 0.020(2) & 4.251(27) & 8.740(53)  & 6.711(134)  & 6.3      \\
V12   & 16.343(3)  & 0.422(4) & 0.219(4) & 0.142(4) & 0.104(5) & 3.999(30) & 8.369(45)  & 6.351(60)   & 1.7      \\
V13   & 16.133(1)  & 0.345(2) & 0.182(2) & 0.114(1) & 0.076(1) & 4.086(12) & 8.438(18)  & 6.608(28)   & 1.3      \\
V22   & 16.166(1)  & 0.226(1) & 0.102(1) & 0.051(1) & 0.022(1) & 4.315(18) & 8.931(31)  & 7.344(66)  & 2.3      \\
V23   & 16.225(1)  & 0.140(1) & 0.041(1) & 0.013(1) & 0.008(1) & 4.442(38) & 9.013(119) & 7.755(233)  & 8.9      \\
V24   & 16.228(2)  & 0.329(2) & 0.180(2) & 0.111(2) & 0.071(1) & 4.156(18) & 8.564(29)  & 6.821(37)   & 1.2      \\

\hline

             &            &           &           & RRc stars          &  &           
 &             &             &\\
\hline

V18 & 16.151(1) & 0.226(2) & 0.012(2) & 0.019(2) & 0.012(2) & 5.775(132)  & 4.629(92)    & 2.901(140)   & - \\
V19 & 16.000(1) & 0.192(2) & 0.014(2) & 0.012(2) & 0.012(2) & 7.1176(125) & 4.836(145)   & 4.159(135)   & - \\
V20 & 16.202(3) & 0.190(3) & 0.011(3) & 0.008(3) & 0.006(3) & 3.156(322)  & 2.119(475)   & 6.918(465)   & - \\
V26 & 16.162(1) & 0.047(2) & 0.007(2) & 0.018(2) & 0.015(1) & 7.7262(260) & 4.7131(1024) & 4.989(1000)  & - \\

\hline	
\end{tabular}
\end{center}
\raggedright
\end{table*}

The Fourier coefficients of the RR Lyrae stars identified in NGC 6712 are listed in Table \ref{tab:fourier_coeffs}, and in Table \ref{fisicos} we report the physical parameters derived from the Fourier decomposition. In their paper, \citetalias{Jurcsik1996} propose a test that validates the usability of the Fourier coefficients obtained from the Fourier decomposition of the light curves of RRab stars. They suggest that the values of the iron abundance [Fe/H] obtained are only valid if their {\it Deviation Parameter} or $Dm$  does not exceeds the value of 3.0. 
The $Dm$ value for each RRab star is listed in Table \ref{tab:fourier_coeffs} in column 10. 

The iron abundance in the scale of \citet{Zinn1984} can be converted into the high-dispersion spectroscopy (HDS) or UVES scale of \citet{Carretta2009} via the equation [Fe/H]$_{\rm UVES} = -0.413 + 0.130$[Fe/H]$_{\rm ZW} -0.356$[Fe/H]$^2_{\rm ZW}$, and are also listed in Table \ref{fisicos}.

\begin{table*}
\footnotesize
\begin{center}
\caption[] {\small Physical parameters obtained from the Fourier fit for the RRab and RRc stars. The numbers in
parentheses indicate the uncertainty on the last decimal place. See $\S$  \ref{decomposition} for a detailed discussion.}
\label{fisicos}
\hspace{0.01cm}
 \begin{tabular}{lllllllll}
\hline 
 &  & & & RRab stars &  & & \\
\hline
Star ID &[Fe/H]$^{\rm JK96}_{\rm ZW}$&[Fe/H]$^{\rm JK96}_{\rm UVES}$ &[Fe/H]$^{\rm N13}_{\rm UVES}$ &$M_V$ & log~$T_{\rm eff}$ & log$(L/{L_{\odot}})$ & $M/{M_{\odot}}$&$R/{R_{\odot}}$\\

\hline
V1        & $-1.23$(4)   & $-1.11$(3)  & $-0.82$(6)  & 0.605(4)  & 3.819(9)  & 1.658(2) & 0.68(7)  & 5.20(1)  \\
V3        & $-1.30$(5)   & $-1.18$(5)  & $-0.77$(10)  & 0.539(3)  & 3.800(20) & 1.685(1) & 0.65(11) & 5.87(1)   \\
V4$^{1}$  & $-1.26$(5)   & $-1.14$(4)  & $-0.80$(8)  & 0.614(3)  & 3.803(20) & 1.654(1) & 0.63(12) & 5.59(1)   \\
V5        & $-1.30$(3)   & $-1.18$(3)  & $-0.94$(5)  & 0.570(4)  & 3.815(9)  & 1.672(2) & 0.69(8)  & 5.41(1)  \\
V6$^{1}$  & $-0.80$(5)   & $-0.74$(3)  & $-0.26$(5)  & 0.662(3)  & 3.823(20) & 1.635(1) & 0.62(12) & 4.98(1)   \\
V12$^{1}$ & $-1.12$(42)  & $-1.00$(34) & $-0.64$(6) & 0.598(6)  & 3.822(10) & 1.661(2) & 0.68(8)  & 5.15(1)  \\
V13       & $-1.28$(2)   & $-1.16$(2)  & $-0.90$(3)  & 0.574(2)  & 3.812(8)  & 1.670(1) & 0.67(6)  & 5.45(1)   \\
V22       & $-1.16$(4)  & $-1.05$(4) & $-0.53$(5)   & 0.535(1)  & 3.803(10) & 1.686(1) & 0.63(7)  & 5.79(1)   \\
V23$^{1}$ & $-1.04$(11)  & $-0.93$(8)  & $-0.39$(16)   & 0.619(1)  & 3.801(27) & 1.652(1) & 0.60(19) & 5.63(1)   \\
V24       & $-1.21$(3)   & $-1.09$(2)  & $-0.78$(4)  & 0.569(3)  & 3.812(8)  & 1.672(1) & 0.66(6)  & 5.48(1)   \\
\hline
Weighted Mean& $-1.25$ & $-1.13$ &$-0.82$ &0.551 & 3.812 & 1.679 & 0.66 & 5.64 \\
$\sigma$ &$\pm$ 0.02   &$\pm$ 0.02 &$\pm$ 0.06 &$\pm$ 0.039  &$\pm$ 0.003  &$\pm$ 0.001 & $\pm$ 0.01 & $\pm$ 0.08 \\
\hline

 &  & & & RRc stars &   & & \\
 \hline

V18       & $-1.17$(23)  & $-1.05$(19) & $-0.97$(10)  & 0.476(11) & 3.866(1) & 1.710(4) & 0.51(1)  & 4.45(2) \\
V19       & $-1.64$(45)  & $-1.58$(52) & $-1.60$(16) & 0.355(10) & 3.857(1) & 1.758(4) & 0.52(1)  & 4.90(2) \\
V20       & $-0.96$(19) & $-0.87$(13) & $-0.87$(7) & 0.650(5) & 3.878(1) & 1.640(2) &0.61(1) & 3.90(1) \\

\hline

\hline
Weighted Mean& $-1.10$ & $-0.95$ &$-0.96$& 0.572 & 3.869 & 1.671 & 0.56 & 4.09\\
$\sigma$ &$\pm$ 0.16 &$\pm$0.17 &$\pm$ 0.19 &$\pm$0.070  &$\pm$ 0.005 & $\pm$ 0.028 & $\pm$ 0.03& $\pm$ 0.24 \\
\hline
\end{tabular}
\end{center}
\raggedright
\center{
1. Not included in the [Fe/H] averages. \\
}
\end{table*}

The values of $M_{V}$ reported in Table \ref{fisicos} have been transformed to luminosities using the following equation:

\begin{equation}
    \rm log(L/L_{\odot}) = -0.4(M_{V} - M^{\odot}_{bol} + BC).
\end{equation}

To calculate the bolometric correction, we used the equation $BC = 0.06[\rm Fe/H]_{ZW} + 0.06$ as derived by \citet{SanCac1990}. We have adopted the value of  $M^{\odot}_{bol}$ = 4.75 mag.

The effective temperature of the RRab stars was estimated with the calibration given by \citet{Jurcsik1998}:

\begin{equation}
    \rm log(T_{\rm eff}) = 3.9291-0.1112(V-K)_{0}-0.0032[\rm Fe/H],
\end{equation}

\noindent
where 

\begin{equation}
(V-K)_{0} = 1.585+1.257P-0.273A_{1} - 0.234\phi_{31}^{(s)} + 0.062\phi_{41}^{(s)}. 
\end{equation}

For the effective temperature in RRc stars, we used:

\begin{equation}
    \rm log(T_{\rm eff}) = 3.7746 - 0.1452\rm log(P) + 0.0056\phi_{31}^{(c)},
\end{equation}

\noindent
as derived by  \citet{Simon1993}. We can also estimate the masses of the RR Lyrae stars using: 
$\rm log(M/M_{\odot}) = 16.907 - 1.47\rm log(P_{F}) + 1.24\rm log(L/L_{\odot}) - 5.12\rm log(T_{\rm eff})$
as given by \citet{vAlbada1971} where P$\rm_{F}$ is the fundamental period. The stellar mean radii can be obtained from $L  = 4\pi R^{2} \sigma T^{4}$. These values are also reported in Table \ref{fisicos}. \\ 

\section{ Further comments on the metallicity of NGC 6712}
\label{decomp_Nemec}

NGC 6712 is a moderately metal-poor OoI-type cluster from which \citet{Harris1996} reports a metallicity of [Fe/H] = --1.02. For our estimation of the metallicity, we did not use the RRab stars V4, V6 and V23 since the value of their $Dm$ parameter is larger than 3.0. For V6, its light curve also shows a mild Blazhko-like amplitude modulation which affects the value of its physical parameters. In the case of V12, we recall that its light curve is incomplete due to its period being nearly half-day and although its Dm value is smaller than 3.0, the uncertainties in its physical parameters are very large, and therefore was not taken into account for the averages. Following the Fourier decomposition method described in $\S$ \ref{decomposition}, we found an average [Fe/H]$^{\rm JK96}_{\rm UVES} =-1.13 \pm 0.02$ given in HDS scale \citep{Carretta2009}. 

It has been recognised that the calibration of \citetalias{Jurcsik1996} produces metallicities too rich by $\sim$0.3 dex for extremely metal poor stars (e.g. \citetalias{Jurcsik1996}, \citet{Smolec2005}). For more metal-rich stars the differences between spectroscopic values and those from the \citetalias{Jurcsik1996} calibration are small \citep{Smolec2005}. The present results offer a good opportunity to test the prediction of \citetalias{Jurcsik1996} calibration for the high metallicity extreme, being NGC 6712 among the most metal-rich globular clusters. 
 \citet{Nemec2013} (N13) proposed new formulations to derive photometric values on the HDS scale for the RRab and RRc stars, using non-linear model between HDS values of [Fe/H] and the Fourier parameter $\phi_{31}$, in a broader metallicity range, particularly at the lower end of [Fe/H]. For completeness we reproduce here their formulations.
 
For the RRab stars,
\begin{equation} 
 {\rm [Fe/H]} = b_0 + b_1~P + b_2 \phi^{(s)}_{31}(Kp) + b_3 \phi^{(s)}_{31}(Kp)~P + b_4~(\phi^{(s)}_{31}(Kp))^2,
\label{eq:Nemecab}
\end{equation}

\noindent
with the coefficient values $ b_0=-8.65 \pm 4.64, b_1=-40.12 \pm 5.18, b_2=5.96 \pm 1.72,  b_3=6.27 \pm 0.96$ and $b_4= -0.72 \pm0.17$, and a rms error of 0.084 dex. The Fourier parameter $\phi^{(s)}_{31}(Kp)$, is calculated from the light curve in the Kepler photometric system, thus, our Fourier parameters from the $V$ light curve listed in Table \ref{tab:fourier_coeffs} were transformed into the Kepler system via the relation: $\phi^{(s)}_{31}(Kp) = \phi^{(s)}_{31} + 0.151$ \citep{Nemec2011} before applying the calibration.
 
 For the RRc stars,
\begin{equation} 
 {\rm [Fe/H]} = b_0 + b_1~P + b_2 \phi^{(c)}_{31} + b_3 \phi^{(c)}_{31}~P + b_4 P + b_5~(\phi^{(c)}_{31})^2,
\label{eq:Nemecc}
\end{equation}

\noindent
with the coefficient values $ b_0=1.70 \pm 0.82, b_1=-15.67 \pm 5.38, b_2=0.20 \pm 0.21,  b_3=-2.41 \pm 0.62$, $b_4=18 \pm 8.70$, and $b_5=0.17 \pm 0.04$, and a rms error of 0.13 dex. The Fourier parameter $\phi^{(c)}_{31}$ is straight that given in Table \ref{tab:fourier_coeffs}. 
 
 We used these formulations to calculate the individual values [Fe/H]$^{\rm N13}_{\rm UVES}$ listed in column 4 of Table \ref{fisicos} and the weighted averages $-0.82 \pm 0.06$ for the RRab stars and $-0.96 \pm 0.19$ for the RRc stars. The weighted average of these two values is  $-0.85 \pm 0.05$, which is in good agreement with the value reported by  \citet{Ferraro1999a} of [Fe/H]$^{\rm N13}_{\rm UVES}$ = -0.88.
These values of the iron-to-hydrogen ratio, in the HDS scale, should be compared to the values of [Fe/H]$^{\rm JK96}_{\rm UVES}$ given in column 3 of Table \ref{fisicos}, i.e. the transformation of the values in the ZW scale, into the HDS or UVES scale established by \citet{Carretta2009}. 
In conclusion, the \citetalias{Nemec2013} calibrations give slightly higher iron-to-hydrogen abundances for this metal poor cluster than the \citetalias{Jurcsik1996} calibration, however, the difference is small and both determinations  agree reasonably well with the value of $-1.0$ dex adopted in the compilation of \citet{Harris1996}. Later we shall return to the metallicity issue when we compare isochrones for a given metallicity with the observed star distribution in the CMD (See $\S$ \ref{CMD}).

\section{On the distance to NGC 6712}
\label{distance_determination}
\subsection{The reddening of NGC 6712}
\label{reddening}

The determination of an accurate value for the interstellar reddening is fundamental in the estimation of a distance to NGC 6712. In his catalogue, \citet{Harris1996} (2010 edition) reports a reddening of $E(B-V) = 0.45$ although there is no consensus in the literature with respect its actual value.   
In order to obtain an independent estimate for the reddening, we followed the method used by \citet{Sturch1966}, according to which, RRab stars have a constant intrinsic colour $(B-V)_0$ near minimum light, between phases 0.5 and 0.8. We also made use of the calibration of $(V-I)_0$ in this range of phases derived by \citet{Guldenschuh2005} as $\overline{(V-I)_{o,min}}$ = 0.58 $\pm$ 0.02 mag. This allowed us to estimate the individual values of $E(V-I)$ for six RRab stars, namely, V4, V5, V13, V22, V23 and V24. We then converted to $E(B-V)$ using the ratio $E(V-I)/E(B-V) = 1.259$ derived by \citet{Schlegel1998}. This yielded a mean reddening $E(B-V) = 0.35 \pm 0.04$ which we have adopted for our calculations. The light curves were phased with the periods listed in Table \ref{member_variables} and the values for the reddening of each star used are in Table \ref{reddening_value}.
Evidences of differential reddening in the region of NGC 6712 were offered by  \citet{Janulis1992}, which can explain the inconsistencies found in the literature for the several reddening estimations, which will impact the determination of the distance to the cluster. See for instance table 3 in \citet{Janulis1992} and a further discussion in $\S$ \ref{distance_inconsistencies}.

\begin{table}
\caption{Reddening estimations from RRab stars.}
\label{reddening_value}
\centering                   
\begin{tabular}{lc}
\hline
Star ID     &  $E(B-V)$  \\
\hline
V4  & 0.409 $\pm$ 0.005 \\
V5  & 0.330 $\pm$ 0.015 \\
V13 & 0.308 $\pm$ 0.007 \\
V22 & 0.335 $\pm$ 0.003 \\
V23 & 0.326 $\pm$ 0.023 \\
V24 & 0.388 $\pm$ 0.008 \\
\hline

\hline	
Mean &  0.349\\
$\sigma$ & $\pm$ 0.036\\
\hline
\end{tabular}
\end{table}

\subsection{From the RR Lyrae stars}
From the Fourier decomposition of the light curves of the RRab and RRc stars and the empirical calibrations for the absolute magnitude $M_V$ given in $\S$  \ref{decomposition}, we found a distance of 8.1 $\pm$ 0.3 kpc and a distance of 8.0 $\pm$ 0.3 kpc, respectively. As an independent estimation for the distance, we also made use of the P-L relation for RR Lyrae stars in the $I$-band \citep{Catelan2004}, $M_I = 0.471-1.132~ {\rm log}~P +0.205~ {\rm log}~Z$, with ${\rm log}~Z = [M/H]-1.765$; $[M/H] = \rm{[Fe/H]} - \rm {log} (0.638~f + 0.362)$ and log~f = [$\alpha$/Fe], from where we adopted [$\alpha$/Fe]=+0.3 \citep{Sal93}. With the aforementioned relation, we derived a distance of 8.3 $\pm$ 0.3 kpc. The value of the extinction coefficient adopted was {\nobreak $A_{v}=3.1E(B-V)$}. We used all the RR Lyrae stars listed in table \ref{tab:fourier_coeffs}.

\subsection{Luminous Red Giants as distance indicators}
A method that was originally developed to determine distances to nearby galaxies \citep{Lee1993} can in principle be used to determine the distance to a globular cluster, using the luminosity of the brightest stars near the tip of the red giant branch (TRGB). The bolometric magnitude of the TRGB can be estimated from the cluster metallicity via the equation \citep{SalCas1997}:

\begin{equation}
\label{TRGB}
M_{bol}^{tip} = -3.949\, -0.178\, [M/H] + 0.008\, [M/H]^2,
\end{equation}

\noindent
where $[M/H] = \rm{[Fe/H]} - \rm {log} (0.638~f + 0.362)$ and log~f = [$\alpha$/Fe]
\citep{Sal93}.

When using this method, one should be aware of two things: first, this method strongly depends on an adequate selection of the stars used, and second, that the brightest stars in a given cluster may not actually be at the tip of the RGB in the CMD, but rather beneath it by a certain magnitude. \citet{Viaux2013}  argued that, in low-mass stars, the neutrino magnetic dipole  moment, may delay the helium ignition and  result in the extension of the red giant branch, and that, in the case of M5 the brightest stars are between 0.04 and 0.16 mag below the TRGB.
The suggested offset was confirmed by the non-canonical models of \citet{Arceo2015}; from the analysis of 25
globular clusters these authors concluded that the theoretical TRGB is in average about $0.26\pm0.24$ bolometric magnitudes brighter than the one observed.
Using eq. \ref{TRGB} and the SR-type star members in the cluster, we estimated its distance and found that if we use a correction of 0.1mag, V8 and V15, which are near the TRGB following the course of the two extreme isochrones in Fig. \ref{CMD_6712}, yield values of 8.0 kpc and 8.2 kpc respectively, which are in good agreement with the estimated distances obtained by the other methods discussed previously. This probably suggests that the true TRGB in NGC 6712 is located at about 0.1mag above these stars. 

\subsection{Cluster distance inconsistencies}
\label{distance_inconsistencies}
The RR Lyrae star light curve Fourier decomposition and the $I$-band P-L relationship lead to a distance of about 8.1$\pm$0.2 kpc (Table \ref{distance}) if a reddening $E(B-V)=0.35$ is assumed ($\S$ \ref{reddening} and Table \ref{reddening_value}). This value is over one kiloparsec larger than 6.9 kpc for $E(B-V)=0.45$ adopted by \citet{Harris1996}, and the kinematic value 6.95$\pm$0.39 kpc obtained by \citet{Baumgardt2019} from the analysis of \textit{Gaia}-DR2 proper motions and radial velocities. We must stress that if $E(B-V)=0.45$ is adopted, our Fourier and P-L values would result in a distance of 7.0$\pm$0.2 kpc. The galactic dust maps and calibrations of \citet{Schlegel1998} and \citet{Schlafly2011} give $E(B-V)$ values of 0.397 $\pm$  0.007 and 0.342 $\pm$  0.006 respectively. 
The 3D Dust Mapping facility mentioned in \ref{ew_stars} suggests the increase trend of $E(B-V)$ with distance in a given direction, and suggests for NGC 6712, $E(B-V)$ $\sim$ 0.37-0.40 at 7.9 kpc. Thus we consider very unlikely that the reddening is above 0.40, in which case the distance yielded by our methods is hardly below 7.5 $\pm$ 0.2 kpc. However, we must add that, if a reddening of 0.40 is taken, the isochrones and ZAHB of the CMD will appear unacceptably shifted to the red (see $\S$ \ref{CMD}). Thus, the RR Lyrae distance and our approaches
lead to a distance, given the respective uncertainties, marginally larger than the kinematic result. Finally, we want to point out to two independent photometric reddening and distance determinations for NGC 6712: $E(B-V)=0.33$ and $7.9\pm1.0$ kpc \citep{Ortolani2000} and $E(B-V)=0.33\pm0.05$ and $\sim$ 8.0 kpc \citep{Paltrinieri2001}. Both values were obtained by considering that the colour of the RGB at the level of the HB of the globular cluster NGC 6712 is a close analogue to NGC 6171. This difference in colour yields  $\Delta(B-V)_{NGC6712-NGC6171}$ = 0.0, meaning that these two have the same reddening.

\begin{table} 
%\scriptsize
\begin{center}
\caption{Distance comparison to NGC 6712 from the different methods used in this work.}
\label{distance}

\begin{tabular}{lc}
\hline
Method & Distance [kpc]        \\
\hline
RRab Fourier decomposition      & 8.1 $\pm$ 0.3 \\
RRc  Fourier decomposition      & 8.0 $\pm$ 0.3   \\    
RRab / RRc $I$-band P-L         & 8.3 $\pm$ 0.3  \\        
\hline
\end{tabular}
\end{center}

\end{table}

\section{The Colour-Magnitude Diagram of NGC 6712}
\label{CMD}

\begin{figure*}
\begin{center}
\includegraphics[width=\textwidth, height=9cm]{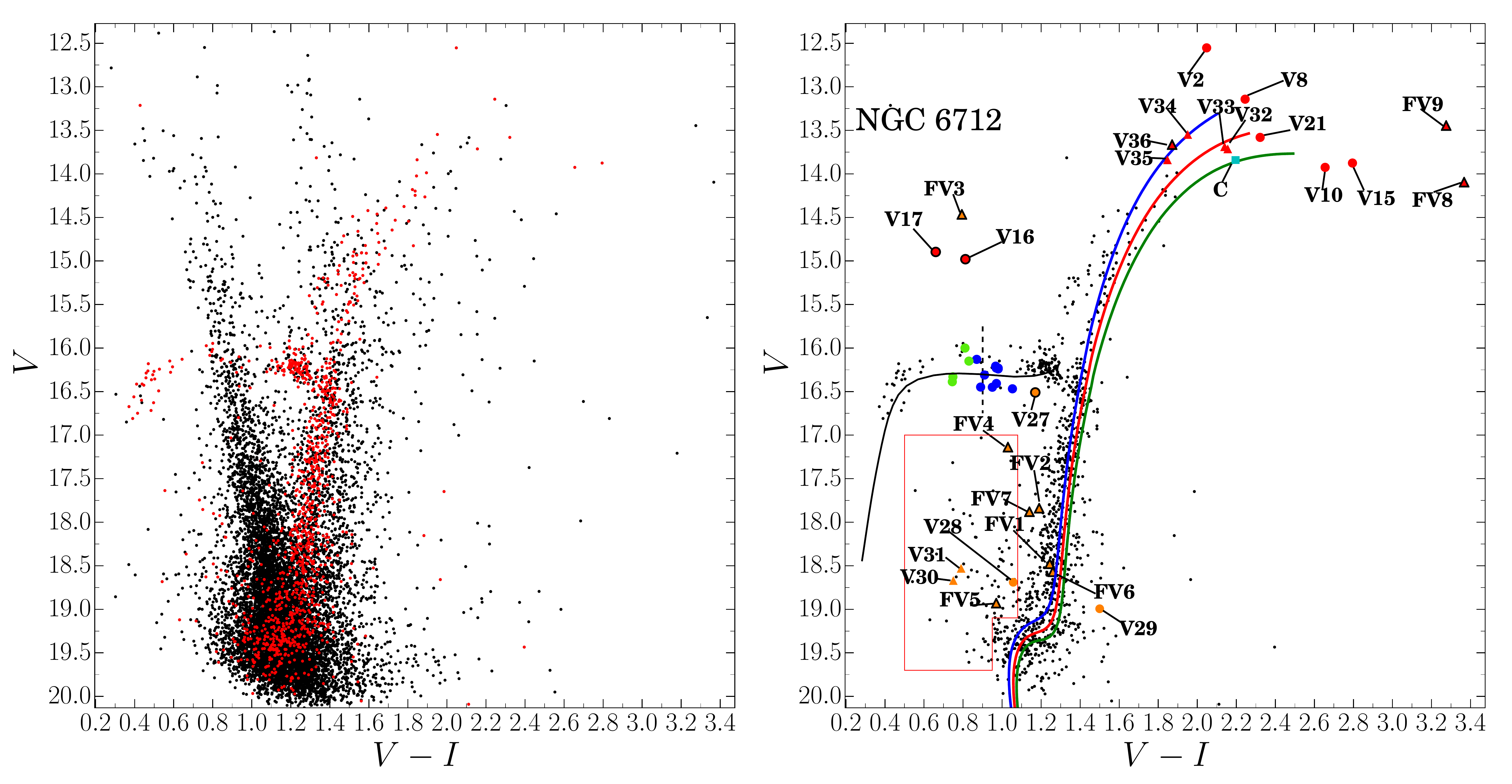}

\includegraphics[width=13cm, height=4cm]{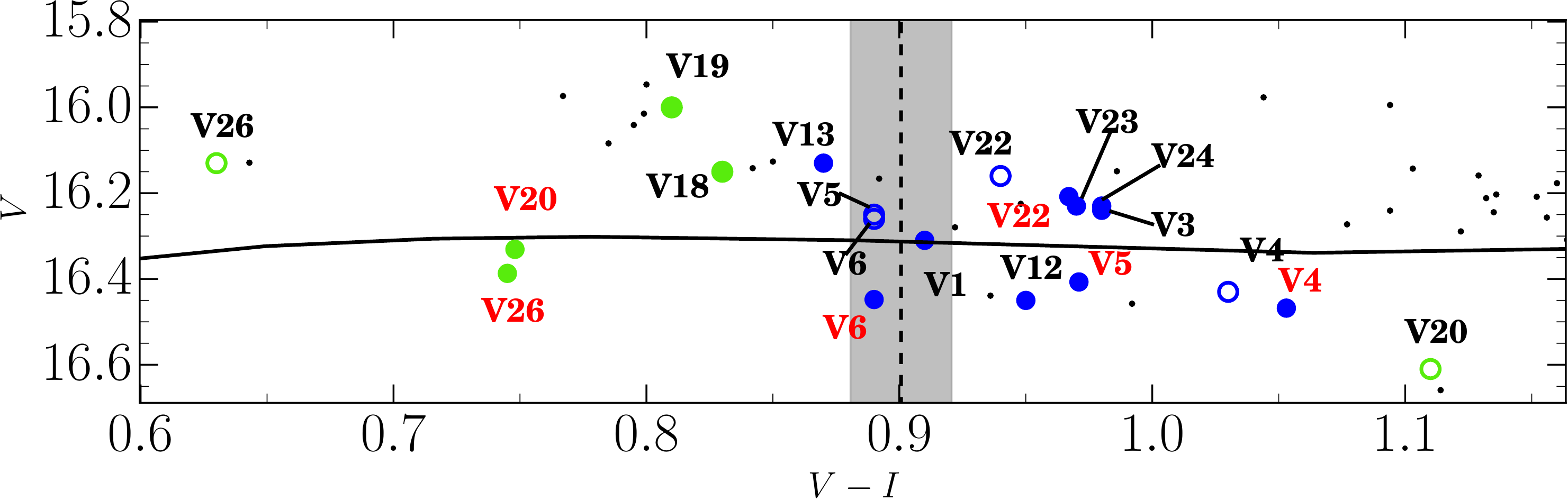}
\caption{Colour-Magnitude Diagrams of NGC 6712 in $\emph{VI}$ filters. The left panel shows the CMD with all the measured stars in our FoV (black dots) and the likely star members (red dots). The right panel shows the likely star members once the CMD has been cleaned of field stars following the method described in $\S$  \ref{membership}. The right panel also shows the known and newly discovered variables. We possess the light curves of 1100 stars (members and non-members). The blue circles correspond to RRab stars, the green circles to RRc stars, the red circles to long-period stars and the orange circles to EW stars. Triangular markers are newly discovered stars. Markers with a black rim represent non-member variable stars. The cyan square represents a candidate variable star. The isochrones in the right panel are 12 Gyr interpolations for [Fe/H] values $-1.25$ (blue), $-1.0$ (red) and $-0.86$ (green). See $\S$ \ref{CMD} for a discussion on the average metallicity. All three isochrones were created using  Y=0.25 and [$\alpha$/Fe]=+0.4, from the model grid of \protect\citet{Vandenberg2014}. The isochrones and the ZAHB (black line) have been shifted to a distance of 8.1 kpc and reddened by $E(B-V)$ = 0.35. The red box delimits the region where blue straggler stars are found (see \ref{BS} for details). The lower panel shows an expansion of the HB for clarity purposes. The open blue and green circles correspond to the original position of the RR Lyrae before their magnitude and colour were corrected due a contaminating neighbouring star. The red tags accompany the new position after the correction. The dashed vertical line represents the FORE of the instability strip as estimated by \protect\citet{Arellano2016} with its uncertainty in grey. }
\label{CMD_6712}
\end{center}
\end{figure*}

Given its position near the Galactic Bulge, NGC 6712 presents a highly contaminated CMD. This can be seen in Fig. \ref{CMD_6712}, left panel.
From their \emph{BV} photometry and constraining to stars within a radius
r < 47'', \citet{Ortolani2000} managed to clean the CMD, but no membership analysis was actually carried out.

With the method described in $\S$  \ref{membership}, we were able to remove the field stars and obtained a remarkably clean CMD, see Fig. \ref{CMD_6712}, right panel. This allowed us to overlay three isochrones, one with a metallicity of [Fe/H] = $-1.25$ (blue), which comes from the the value obtained from the Fourier decomposition of the light curves of the RR Lyrae stars, another with the metallicity of [Fe/H] = $-1.0$ (red) reported by \citet{Harris1996} and one with [Fe/H] = --0.86 (green) derived from the equations of \citetalias{Nemec2013}, given in $\S$  \ref{decomp_Nemec}. The three isochrones were shifted to a distance of 8.1 kpc and have been reddened by $E(B-V)= 0.35$. The isochrones were calculated from the models of \citet{Vandenberg2014} with Y = 0.25 and [$\alpha$/Fe] = +0.4, and they correspond to an age of 12 Gyrs. It is evident that the two isochrones with extreme metallicities, bracket the one for [Fe/H]= $-1.0$, which, despite the scatter at the RGB, particularly near the tip, seems to produce the best fit. In other words, the observed CMD supports the conclusion that [Fe/H]= $-1.0$ is indeed a solid estimation. The iron-to-hydrogen calibrations of \citetalias{Jurcsik1996} and of \citetalias{Nemec2013}, discussed in sections \ref{decomposition} and \ref{decomp_Nemec} respectively, agree, within their own uncertainties, with this value of the metallicity.

Like the isochrones, a ZAHB of metallicity [Fe/H]=-1.31 \citep{Vandenberg2014} was placed in the CMD. No significant differences were observed for other metallicities. The resulting position of the ZAHB is consistent with the distance of 8.1 kpc.

\subsection{The structure of the Horizontal Branch of NGC 6712}

Once the CMD has been cleaned from field stars using the method described in $\S$  \ref{membership}, it displays a prominent red HB and a sparsely populated blue tail  which is typical of OoI-type clusters. The RR Lyrae population on the HB is dominated by fundamental mode RRab pulsators (10) in comparison with first overtone pulsators (4). The distribution of RRab and RRc shows a clear mode segregation around the First Overtone Red Edge (FORE) if one accepts that the FORE may be shifted to the blue by a few hundredths of a magnitude, of the otherwise established position (see Fig. \ref{CMD_6712}), which given the uncertainties, it is quite likely. 
We can also see that V22 (RR Lyrae with no previous classification) and V23 (originally classified as RRc) clearly fall on the right side of the FORE of the IS which along with their period, light curve shape, and position in the Period-Amplitude diagram (Fig. \ref{baileyDiagram}) confirm their nature as RRab stars.

While plotting all the positions of the \textit{Gaia} sources in our FoV, we noticed that two sources, not resolved in our photometry, fall within the FWHM of the PSF of some of the RR Lyrae stars, namely V4, V5, V6, V20, V22 and V26. The data of the two Gaia sources for each variable are listed in Table \ref{variables_gaia}. Hence, their magnitude and colour are contaminated by the extra flux of the neighbour. A correction to their positions on the HB was performed by measuring the combined flux in the $G, G_{BP}-$, and $G_{RP}$-band, of both sources, and converting them to the Johnson $V_{GAIA}$ and $I_{GAIA}$ magnitudes. For this purpose we used the relations provided by J.M. Carrasco (2018: Gaia team), available in Section 5.3.7 of the \textit{Gaia}-DR2 documentation \footnote{http://gea.esac.esa.int/archive/documentation/GDR2/index.html}. The magnitudes of the two sources were combined to calculate $V_{mix}$ and $I_{mix}$,  and the corrections $\Delta V= V_{GAIA}-V_{mix}$ and $\Delta I= I_{GAIA}-I_{mix}$ where then added to the intensity-weighted means $\big<V\big>$ and $\big<I\big>$ (Table \ref{member_variables}) calculated from our light curves, i.e., to estimate their apparent magnitude without the flux contamination of the neighbouring star. This also helped us to recalculate the amplitude for these stars and, in doing so, their position in the Bailey diagram \ref{baileyDiagram}. In the lower panel of Fig. \ref{CMD_6712} we plot the corrected positions of the aforementioned RR Lyrae stars.

\begin{table*}
%\scriptsize
\begin{center}
\caption{Data of RR Lyrae stars (in bold face) in the field of NGC 6712, which show flux contamination from an adjacent star (within the PSF) from \textit{Gaia}-DR2.}
\label{variables_gaia}

\begin{tabular}{llccccccccccc}
\hline
 
Variable & \textit{Gaia}-DR2 ID & $G$ & $G_{BP}-G_{RP}$ & $V_{Gaia}$ & $I_{Gaia}$ & $V-I_{Gaia}$ & V$_{mix}$ & $\Delta_{V}$ & $I_{mix}$ & $\Delta_{I}$ & Amp$_{V}$ & Amp$_I$ \\
 \hline

\textbf{V4}  & \textbf{4203849122517862144} & \textbf{16.154} &  \textbf{1.239} & \textbf{16.446} & \textbf{15.362} &  \textbf{1.084} & \textbf{16.408} &  \textbf{0.038} & \textbf{15.347} & \textbf{0.015} & \textbf{0.58} & \textbf{0.35}   \\
V4  & 4203849122443611648 & 20.053 &        & 20.070 & 20.032 &  0.038 &        &        &        &       &       &         \\
\textbf{V5}  & \textbf{4203848778920519168} & \textbf{15.911} &  \textbf{1.005} & \textbf{16.110} & \textbf{15.242} &  \textbf{0.869} & \textbf{15.954} &  \textbf{0.157} & \textbf{15.166} & \textbf{0.076} & \textbf{1.34} & \textbf{0.81}   \\
V5  & 4203848778846478080 & 18.114 &        & 18.132 & 18.093 &  0.038 &        &        &        &       &       &         \\
\textbf{V6}  & \textbf{4203848985005261312} & \textbf{16.147} &        & \textbf{16.164} & \textbf{16.126} &  \textbf{0.038} & \textbf{15.976} &  \textbf{0.188} & \textbf{15.938} & \textbf{0.188} & \textbf{1.06} & \textbf{0.66}   \\
V6  & 4203848985005260672 & 17.954 &        & 17.972 & 17.934 &  0.038 &        &        &        &       &       &         \\
\textbf{V20} & \textbf{4203849083790434304} & \textbf{16.042} &  \textbf{0.999} & \textbf{16.240} & \textbf{15.377} &  \textbf{0.863} & \textbf{16.069} & \textbf{0.171} & \textbf{15.293}  & \textbf{0.083} & \textbf{0.48}      & \textbf{0.26}        \\
V20 & 4203849088158325760 & 18.143 &        & 18.161 & 18.123 &  0.038 &        &       &         &       &       &         \\
\textbf{V22} & \textbf{4203849874064518016} & \textbf{15.945} &  \textbf{1.115} & \textbf{16.186} & \textbf{15.216} &  \textbf{0.969} & \textbf{16.138} & \textbf{0.048} & \textbf{15.196} & \textbf{0.021}  & \textbf{0.61} & \textbf{0.40}   \\
V22 & 4203849878432420352 & 19.528 &        & 19.545 & 19.507 &  0.038 &        &       &        &        &       &         \\
\textbf{V26} & \textbf{4203849466042002944} & \textbf{15.980} & \textbf{0.861}  & \textbf{16.132} & \textbf{15.392} &  \textbf{0.740} & \textbf{15.874} & \textbf{0.257} & \textbf{15.249} & \textbf{0.142}  & \textbf{0.15} & \textbf{0.10}   \\
V26 & 4203849466042002816 & 17.545 &        & 17.563 & 17.525 &  0.038 &        &       &        &        &       &         \\
\hline
\end{tabular}
\end{center}
\end{table*}

At this stage, we can comment on the RR Lyrae pulsating-mode distribution and their relation with the HB overall structure.
A parameter that is useful in describing the morphology of the HB is the Lee parameter \citep{Lee1990} defined as $\mathcal{L} = (B-R)/(B+V+R)$ where $B$, $V$ and $R$ are the number of stars to the blue, inside and to the red of the IS. For NGC 6712, we estimated from the clean CMD of Fig. \ref{CMD_6712}, $\mathcal{L}$ = $-0.44$. For the sake of comparison, \citet{Paltrinieri2001} reported $\mathcal{L}$ = $-0.66$, although they estimated this value by constraining the stars in their FoV to a radius $<$5' so it might still include some foreground stars on the HB which would make up for the difference with respect to our estimated value. 
It has been highlighted in several recent studies, e.g. \citet{Arellano2019} (their figure 8), that the RRab and RRc stars may be cleanly separated by the FORE or that a mixture of modes can be found in the bi-modal region, also sometimes referred to as the either-or region. And, that this condition seems to be correlated with the overall structure parameter $\mathcal{L}$ and the Oosterhoff type. There are six OoI-type clusters with very red HB's ($\mathcal{L}$ < $-0.5$), Rup 106, NGC~362, NGC~1261, NGC~6171, NGC~6362 and NGC~6712; for NGC~362 there is no detailed study of the mode distribution and Rup 106 has no known RRc stars.
In the bottom panel of Fig. \ref{CMD_6712} we note that V13 might be the only RRab star sitting in the bimodal region. However, given the uncertainties in the stellar colours and the positioning of the FORE the two modes could be considered as being segregated, in which case, in NGC 6712, like in the other three, NGC~1261, NGC~6171 and NGC~6362, the modes are well separated. This property is also observed in all OoII-type clusters thus far studied but only in some OoI-type clusters of intermediate values for $\mathcal{L}$.

\subsection{Comments on the blue-straggler population of NGC 6712}
\label{BS}
\citet{Paltrinieri2001}, based on their $V-(B-V)$ CMD,  identified 108 candidate blue-straggler stars (BSS) in the central region of NGC 6712. The cluster being a very compact object, finding a large number of BSS does not comes as a surprise, since they may originate in stellar collisions \citep{Ferraro1999}. The BSS region in our CMD does not look so much populated. We identified 61 cluster members in this region after cleaning for the non-members through the proper motion analysis ($\S$ \ref{membership}). We also found two contact binaries among these BSS, V30 and V31. Thus, we are inclined to believe that NGC 6712 does not have a large population of BSS as previously suggested by \citet{Paltrinieri2001}

\section{COMMENTS ON THE  NGC 6712 CLUSTER TIDAL IDENTITY}
\label{dynamics}

As we have briefly summarised in the introduction, there are numerous evidences that favour the idea that NGC 6712 was once was a very massive globular cluster in the Galaxy, and that after numerous close encounters with the Galactic bulge, and the constant interaction with the disc, it has been tidally stripped of nearly 99\% of its mass 
(\citealt{Andreuzzi2001}; \citealt{Paltrinieri2001}). To what extent do these tidal disrupting forces imprint their evidence in the remaining cluster core? The unprecedented accurate proper motions in the \textit{Gaia}-DR2 are the best observations available at present to investigate this question. The proper motion vectors, projected on the plane of the sky, of the 1529 cluster member stars identified in $\S$  \ref{membership} are shown in the bottom panel of Fig. \ref{vpd}, where it is evident that the whole cluster moves as a compact entity, and the stars in solidarity maintain their identity. This is contrary to what is seen for instance in Pal 13 \citep{Yepez2019} (their figure 12), a rather loose cluster being tidally disrupted. 

However, we note in the bottom panel of Fig. \ref{vpd}, that some apparent field variables (labelled in the figure) have proper motions that are consistently nearly perpendicular to the cluster motion, parallel to the Galactic longitude and towards the Galactic centre. This suggests that these stars participate in the local Galactic rotation, as expected in field stars, and not in the present orbital motion of the cluster.

It is also fair to say that based solely on the proper motions of the stars in our FoV, we are unable to tell the difference between the ones belonging to the field and the ones that are actually being or starting to be stripped away from the cluster.

The cluster, in its current stage within its tidal radius, keeps itself rather compact. This allows it to better withstand its gravitational interaction with the Galactic bulge and the Galactic plane now than in the past, when it was much more massive and extended. It is possible that its compactness allows the cluster to preserve its tidal identity for many more encounters to come.

\section{Summary and conclusions}
\label{conclusions}

In this paper we have performed high-precision $VI$ CCD photometry of 11294 stars in the FoV of globular cluster NGC 6712. Our main goal was to obtain physical parameters of the RR Lyrae star population via Fourier decomposition of their light curves. This method yielded metallicites of [Fe/H]$_{\rm zw}$=--1.25 $\pm$ 0.02 and [Fe/H]$_{\rm zw}$=--1.10 $\pm$ 0.16 for the RRab and RRc stars respectively with a weighted mean of [Fe/H]$_{\rm zw}$=--1.23 $\pm$ 0.02 using the calibrations by \citet{Jurcsik1996} and \citet{Morgan2007}. For comparison we estimated the metallicities using the calibration derived by \citet{Nemec2013} and \citet{Nemec2011} which are given in the HDS scale [Fe/H]$^{\rm N13}_{\rm UVES}$ = -0.82 $\pm$ 0.06 for the RRab stars and [Fe/H]$^{\rm N13}_{\rm UVES}$ = -0.96 $\pm$ 0.19 for RRc stars yielding a weighted mean of [Fe/H]$^{\rm N13}_{\rm UVES}$ = --0.85 $\pm$ 0.05 . The
 above values are, within the uncertainties, in good agreement with the value reported by 
 \citet{Harris1996} of [Fe/H]=--1.02. 

NGC 6712 is currently located inside the bulge of the Galaxy and therefore, it is highly contaminated by field stars. A proper motion analysis of 60,447 stars in our FoV was carried out in order to determine membership status and yielded 1529 likely members or 2.5$\%$, for which we possess the light curves of 1100. This allowed us to plot a remarkably clean CMD and to overlay three theoretical isochrones using the models of \citet{Vandenberg2014} and the metallicity found from the Fourier decomposition of the light curves of the RR Lyrae stars. We found that these isochrones are consistent with an age of 12 Gyrs. We also used a theoretical ZAHB with similar metallicity to the isochrones to fit the observed HB. The cleaning of the CMD also allowed us to decrease the estimated number of BSS members in NGC 6712 down to 61, since after the membership analysis many of them turned out to be field stars. 
The distance derived by the methods mentioned in $\S$ \ref{distance_determination} yielded a value of $\big<d\big>$ = 8.1 $\pm$ 0.2 kpc, if a colour excess $E(B-V)=0.35$ is assumed. Evidences have been discussed that this value could be as large as 0.40, in which case the same methods yield a distance of 7.5$\pm$0.2 kpc.

Based on the their light curve morphology, their position on the Bailey diagram and the period analyses of the stars V22, V23 and V27 we were able to classify properly V22 as a RRab star, to reclassify V23 as a RRab and V27 as a EW star. Furthermore, V22 and V23 are both on the right side of the FORE, where only RRab stars are expected to be found.

From the average of the periods of the RRab stars in the cluster ($\big<P\big> = 0.58$), and the distribution of RRab and RRc star in the period-amplitude diagram, we confirm that NGC 6712 is a OoI-type cluster. 
We detected a mild Blazhko effect in the amplitude of V6.
We additionally found nine new EW variables (V30 and V31 and FV1-FV7) and seven SR-type variables (V32--V36 and FV8--FV9). For five of these SR-type (V32,V36, FV8, FV9 and a candidate C), we were able to derive periods ranging between 12-31 days.

There are numerous evidences in the literature  that NGC 6712 is a remnant of a much more massive cluster that has been tidally disrupted by its numerous encounters with the Galactic bulge. We were able, from the proper motion analysis, to find evidence that the present cluster is in fact a compact core that seems to retain its identity in spite of the gravitational interaction with the Galactic bulge and disc. The proper motion of some of the field variable stars are parallel to the galactic longitude and point towards the Galactic center, suggesting that they are field stars participating in the Galactic rotation, not necessarily being gravitationally stripped away. As of now, we do not have a method to properly differentiate star members that are being currently stripped away from the cluster from the ones that belong to the field. It is possible that what we are seeing now is nothing more than the remaining core.

%\newpage
\section*{Acknowledgements}

We are grateful to Dr. Ra\'ul Michel Murillo for kindly providing the standard stars in the field of NGC 6712. We thank the staff of IAO, Hanle and CREST, Hosakote, that made these observations possible. The facilities at IAO and CREST are operated by
the Indian Institute of Astrophysics, Bangalore. This project was partially supported by DGAPA-UNAM (Mexico) via grant IN106615-17. DD thanks CONACyT for the PhD scholarship. We have made extensive use of the SIMBAD and ADS services.

%%%%%%%%%%%%%%%%%%%%%%%%%%%%%%%%%%%%%%%%%%%%%%%%%%



%\bibliographystyle{mnras}

%\newpage
\bibliographystyle{mnras}
\bibliography{6712_MAIN}

%%%%%%%%%%%%%%%%%%%%%%%%%%%%%%%%%%%%%%%%%%%%%%%%%%

%%%%%%%%%%%%%%%%% APPENDICES %%%%%%%%%%%%%%%%%%%%%

\appendix
\label{appendix}

\section{Field variable and peculiar stars}

\subsection{The case of V7}

This variable was discovered by \citet{Sawyer1953} and classified as a Mira-type star by \citet{Sloan2010}, who were able to derive a pulsation period of 193.0 days. Our data span only 30 days but it displays a clear variation (see Fig. \ref{high-amp}). These are evolved stars that during their evolution on the AGB can produce and expel dust into the interstellar medium, affecting their local reddening. In the case of V7, its location on the CMD is extremely shifted to the red probably due to its own dust shrouding. Hence, it is not labelled in Fig. \ref{CMD_6712} due to scale reasons. Its CMD coordinates are (4.701,16.98). According to the membership analysis carried out in $\S$  \ref{membership}, the star is a cluster member. 

\subsection{The case of V14, V16 and V17}
\label{sr_variables}
None of these three stars is a cluster member according to the analysis described in $\S$  \ref{membership}, in agreement with the proper motion analysis performed by \citet{Cudworth1988} for V17. The SR star V14 is not in the FoV of our images but in terms of variability, V14 is a clear variable \citep{Rosino1966}.  V16 and V17 were flagged as variables by \citet{Harwood1960},  but \citet{Rosino1966} and  \citet{Sandage1966} did not detect any variability. Unfortunately the light curves of V16 and V17 were not published by previous authors. The light curves in our data are shown in Fig. \ref{high-amp}, where subtle but clear signs of long-term variations can be seen.

\begin{figure*} 
\begin{center}
\includegraphics[width=17.0cm,height=8.0cm]{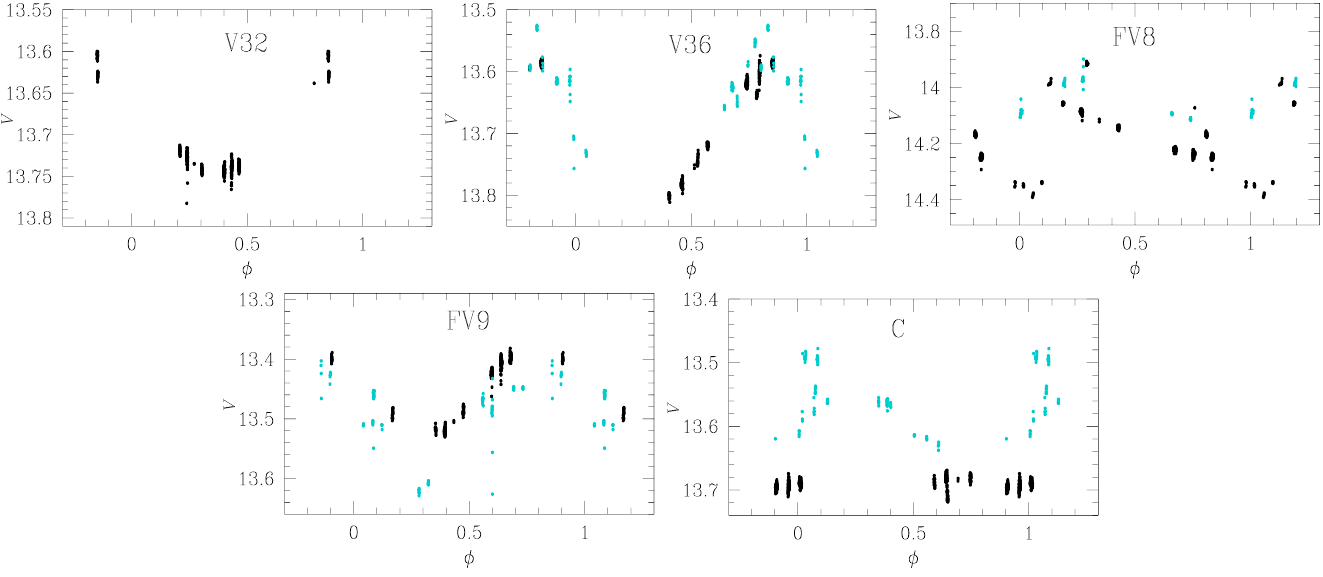}
\caption{SR stars that could be phased with a single period, given in Tables \ref{member_variables} and \ref{field_variables}.}
\label{SR_FASES}
\end{center}
    
\end{figure*}

\subsection{The case of V22}
\label{V22}
\citet{Cudworth1988} noted that this star showed low amplitude variations and that given its position on the CMD, it could presumably be catalogued as a RR Lyrae. In the Catalogue of Variable Stars in Globular Clusters (CVSGC) \citet{Clement2001} the star is designated as RR type and its period is not provided. V22 is also located near the locus where RRab stars in OoI-type clusters are expected to be found in the Period-Amplitude diagram (Fig. \ref{bailey}). The shape of the light curve (see Fig. \ref{grid_rr}), its amplitude, position on the Horizontal Branch (HB) (Fig. \ref{CMD_6712}) and period ($P=0.654789d$), clearly indicate its RRab nature.   

\subsection{The case of V23}
\label{v23}

This star, discovered by \citet{Tuairisg2003} (P1 in their paper), has been classified as RRc. Nevertheless, its reported period in the CVSGC of $P=0.3896d$ does not phase the curve well. Moreover, using our 7 year time-base data, we found a period of $P=0.642451d$ which is more consistent with a RRab star. Also, like V22, its position on the HB (Fig. \ref{CMD_6712}) and on the Bailey diagram (Fig. \ref{bailey}) along with the shape of its light curve (Fig. \ref{grid_rr}) are all consistent with the star being a RRab.

\subsection{The case of V25}

This is a low-mass X-ray binary discovered by \citet{Homer1996}. According to these authors the optical counterpart shows variability and has a mean magnitude $V=20.34\pm0.07$, i.e. too faint for our CCD photometry. The nearest optical source in our images, to the coordinates listed by the CVSGC, is a $V\sim 16$ mag star which shows no variations.

\subsection{The case of V27}
\label{V27}
This star was discovered and classified as RRc  by \citet{Pietrukowicz2004} (NGC6712\_07 in their paper). These authors noted that the star showed an "unstable light curve with some cycle-to-cycle changes". We found that once it is phased with a period of $P=0.425714d$ (which is approximately twice the reported value by \citet{Pietrukowicz2004}) the light curve of V27 (Fig. \ref{grid_rr}) shows alternating deep and shallow minima typical of EW stars. Hence, the star should be considered a short-period contact eclipsing binary. 

\subsection{Semi-regular red variables}
\label{SPSR}
For five stars among the SR variables, a period that seems to phase all the available data was identified and are given in Tables \ref{member_variables} and \ref{field_variables}. These stars and their phased light curves are shown in Fig. \ref{SR_FASES}. Their amplitudes are of a few tenths of a magnitude and their periods range between 12 and 31 days. This suggests a classification as of the semiregular or SRs type.

%%%%%%%%%%%%%%%%%%%%%%%%%%%%%%%%%%%%%%%%%%%%%%%%%%

% Don't change these lines
\bsp	% typesetting comment
\label{lastpage}
\end{document}